\documentclass{lmcs}
\usepackage{underscore}  
\newif\ifexperiments
\experimentstrue
\usepackage[ruled]{algorithm2e}
\newtheorem{definition}{Definition}

\usepackage{amsmath}
\usepackage{amssymb}
\usepackage{amsfonts}
\usepackage{tikz}  
\usetikzlibrary{arrows,shapes,automata,backgrounds,petri,arrows.meta,calc,positioning,decorations,backgrounds}

\tikzstyle{tplace}=[circle,draw,inner sep=1.5mm]
\tikzstyle{transition} = [rectangle, draw, inner sep=0,text width=0.4cm, fill=gray!40, minimum height=0.4cm, text centered]
\tikzstyle{state} = [circle, draw, inner sep=0,text width=0.4cm, minimum height=0.4cm, text centered]
\tikzstyle{event} = [rectangle, draw=green!50!black!75,fill=green!50!black!20, inner sep=0,text width=0.7cm,
minimum height=0.5cm, text centered,outer sep=0]
\tikzstyle{cond} = [circle, draw=blue!75,fill=blue!10, inner sep=0,text width=0.4cm, minimum height=0.4cm, text centered]
\tikzstyle{pcond} = [circle, draw=magenta!75,fill=magenta!10, inner sep=0,text width=0.4cm, minimum height=0.6cm, text centered]
\tikzstyle{none} = [draw=none, fill=none, inner sep=0,minimum height=0.2cm]
 \tikzstyle{leer} = [rectangle, draw=none, inner sep=0, text  centered]
\tikzstyle{every label}=[red]
\tikzstyle{edge} = [thick]
\newcommand{\define}{\stackrel{\triangle}{=}}

\newtheorem{Lemma}{Lemma}
\newtheorem{theorem}{Theorem}

\newcommand{\unfolding}{\mathcal{U}}

\newcommand{\markn}{\mathit{M}}
\newcommand\marking[1]{\mathit{Mark}(#1)}
\newcommand\markings{\mathcal{M}}

\newcommand\bad{\mathcal{B}}
\newcommand\badstates{{\markings_\bad}}

\newcommand\kringel[1]{\mathsf{loop}(#1)}
\newcommand\minkringel[1]{\mathsf{mloop}(#1)}
\newcommand\goodstates{{\markings_\mathcal{G}}}

\newcommand\badconfs{{\configs_\mathcal{B}}}

\newcommand\atts{\mathcal{A}}
\newcommand\attn{\mathbf{A}}
\newcommand\basin[1]{\mathcal{B}(#1)}
\newcommand\cbasin[1]{{\mathcal{B}^*}(#1)}
\newcommand{\preset}[1]{{^{\bullet}}{#1}}
\newcommand{\postset}[1]{{{#1}^{\bullet}}}
\newcommand{\adequate}{\prec}

\newcommand{\concurrent}{~\mathbf{co}~}

\newcommand{\conflict}{\mathrel{\#}}        
  
\newcommand{\dircf}{\conflict_\delta}  
\newcommand{\strcf}{\conflict_\sigma}  
\newcommand\cone[1]{[#1]}

\newcommand{\tup}[1]{\langle {#1} \rangle}
\newcommand\onlynetn{{\mathit{N}}}
\newcommand{\net}{\onlynetn}

\newcommand\netn{\mathcal{N}}
\newcommand\petrinet{\netn}
\newcommand{\places}{\mathit{P}}
\newcommand{\place}{\places}
\newcommand\transn{\mathit{t}}
\newcommand\onet{\mathcal{O}}
\newcommand\trans{{\mathit{T}}}
\newcommand\placen{\mathit{p}}
\newcommand\flow{\mathit{F}}
\newcommand\oflow{\mathit{G}}

\newcommand\isom{\mathbf{I}}

\newcommand{\edges}{\mathcal{E}}

\newcommand\prefixn{\Pi}

\newcommand{\prefn}{\Pi}
\newcommand\bys{\begin{eqnarray*}}
\newcommand\eys{\end{eqnarray*}}
\newcommand\bdf{\begin{Definition}}
\newcommand\edf{\end{Definition}}
\newcommand\bet{\begin{thm}}
\newcommand\ent{\end{thm}}
\newcommand\bel{\begin{Lemma}}
\newcommand\bec{\begin{cor}}
\newcommand\enc{\end{cor}}
\newcommand\bum{\begin{enumerate}}
\newcommand\eum{\end{enumerate}}
\newcommand\bit{\begin{itemize}}
\newcommand\eit{\end{itemize}}
\newcommand\bepr{{\bf Proof:} \ \nopagebreak}
\newcommand\eepr{\qedhere \hfill $\square$}

\newcommand\NN{\mathbb{N}}

\newcommand\one{\mathbf{1}}
\newcommand\oplace{\mathit{B}}
\newcommand\otrans{\mathit{E}}
\newcommand\oplacen{\mathit{b}} 
\newcommand\otransn{\mathit{e}}

\tikzstyle{grayzone} =
[rectangle, rounded corners,fill=gray, text=red, minimum width=1.7cm, minimum height=1.1cm, fill opacity = .5
]
\tikzstyle{grayzone2} =
[rectangle, rounded corners,fill=gray, text=red, minimum width=3.1cm, minimum height=2.5cm, fill opacity = .2, draw=white]

\tikzstyle{inner} =
[rectangle, rounded corners,fill=gray, text=red, minimum width=1.5cm, minimum height=3.7cm, fill opacity = .5
]
\tikzstyle{outer} =
[rectangle, rounded corners,fill=gray, text=red, minimum width=4.9cm, minimum height=4.2cm, fill opacity = .2, draw=white]
\tikzstyle{outer2} =
[rectangle, rounded corners,fill=gray, text=red, minimum width=4.9cm, minimum height=5.9cm, fill opacity = .2, draw=white]

\newcommand\infigs{\configs^{\infty}}
\newcommand\hinfigs{\hat{\configs}^{\infty}}
\newcommand\runn{\omega}
\newcommand\runs{\Omega}
\newcommand{\states}{\mathcal{M}}
\newcommand\move[1]{\stackrel{#1}{\rightarrow}}
\newcommand\setgdw{\stackrel{\triangle}{\Longleftrightarrow}}

\newcommand\omove[1]{\stackrel{#1}{\leadsto }}

\newcommand\reach{\mathbf{R}}

\newcommand\confign{\mathit{C}}

\newcommand\configs{\mathcal{C}}

\newcommand\finfigs{\mathcal\configs^{\mathbf{f}}}
 
 \newcommand\doofigs{\mathcal{D}}

 \newcommand\viables{\mathcal{F}}
 \newcommand\free{\mathcal{F}}

\newcommand\mindofigs{\check{\doofigs}}
 \newcommand\mindooout{\mathfrak{D}}

\newcommand\cutn{\mathbf{c}}
\newcommand\cut{\mathbf{cut}}

\newcommand\trunk[1]{\langle#1\rangle}
\newcommand\crest[1]{\mathbf{crest}(#1)}
\newcommand\shaved[1]{\mathbf{shave}(#1)}

\newcommand{\dheight}[1]{\mathbf{dech}(#1)}
\newcommand{\schutz}[1]{\mathbf{prot}(#1)}

\newcommand\watershed{\gamma}
\newcommand\tips{\Gamma}
\newcommand\ridge{\chi}

\newcommand{\worklist}{\mathsf{wl}}
\newcommand\addit{\mathsf{add}}
\newcommand\true{\mathsf{true}}
\newcommand\false{\mathsf{false}}
\newcommand\myleq{\leqslant}

\newcommand\morph{\phi}
\newcommand\fold{\pi}

\usepackage{mathtools}
\usepackage[draft]{fixme}
\usepackage{epsfig}

\usepackage{amsfonts}
\usepackage{amsmath}
\usepackage{hyperref}
\usepackage{graphicx}  
\usepackage{caption}
\usepackage{subcaption}

\SetKwComment{Comment}{/* }{ */}
\def\titlerunning{Attractor basins}

\title{Attractor Basins in Concurrent Systems}
\author[G.K.~Aguirre]{Giann Karlo Aguirre-Sambon\'i\lmcsorcid{0000-0002-3526-7253}}[a,e]
\author[S.~Haar]{Stefan Haar\lmcsorcid{0000-0002-1892-2703}}[d]
\author[L.~Paulev\'e]{Lo\"ic Paulev\'e\lmcsorcid{0000-0002-7219-2027}}[b]
\author[S.~Schwoon]{Stefan Schwoon\lmcsorcid{0000-0001-6622-6510}}[a] 
\author[N.~W\"urdemann]{Nick W\"urdemann\lmcsorcid{0000-0001-7934-820X}}[c]
\address{Universit\'e Paris-Saclay, INRIA and LMF, CNRS and ENS Paris-Saclay, Gif-sur-Yvette, France} \email{(giann-karlo.aguirre-samboni,stefan.schwoon)@inria.fr}
\address{Univ. Bordeaux, Bordeaux INP, CNRS, LaBRI, UMR5800, Talence, France}
\email{loic.pauleve@labri.fr}
\address{Department of Computing Science, University of Oldenburg, Oldenburg, Germany}
\email{wuerdemann@informatik.uni-oldenburg.de}
\address{INRIA Saclay center, \textsc{Musca} team}
\email{stefan.haar@inria.fr}
\address{Mines Paris, PSL Research University, Centre for Computational Biology, 75006 Paris, France \newline Institut Curie, PSL Research University, 75005 Paris, France \newline INSERM, U900, 75005 Paris, France}
\email{giann-karlo.aguirre\_samboni@minesparis.psl.eu}
\begin{document}
\begin{abstract}
A crucial question in analyzing a concurrent system is to determine its long-run behaviour, and in particular, whether there are irreversible choices in its evolution, leading into parts of the reachability space from which there is no return to other parts. Casting this problem in the unifying framework of safe Petri nets, our previous work~\cite{CHJPS-cmsb14} has provided techniques for identifying \emph{attractors}, i.e.\ terminal strongly connected components of the reachability space. What we aim at is to determine the \emph{attraction basins} associated to those attractors; that is, those states from where all infinite runs are doomed to end in the given attractor, as opposed to those that are \emph{free} to evolve differently. Here, we provide a solution for the case of safe Petri nets.
Our algorithm uses net unfoldings and provides a map of all of those  configurations  (concurrent executions of the system)  that lead onto \emph{cliff-edges}, i.e.  any maximal extension for those configurations lies in some basin that is considered fatal. 
\end{abstract}

\maketitle

\section{Introduction}
  With the growing interest in formal methods for biology, the key feature of  \emph{multistability} of systems~\cite{ta90,Plahte1995,Ozbudak2004,Pisarchik2014} comes into focus. It has been studied in other
qualitative models such as Boolean and multivalued networks~\cite{t80,tt95,Richard2019}.
Multistability characterizes many fundamental biological processes, such as cellular differentiation, cellular reprogramming, and cell-fate decision; in fact, stabilization of a cell regulatory network corresponds reaching one of the - possibly many - phenotypes of the cell, thus explaining the important role of multistability in cell biology. However, multistability emerges also in  many other branches of the life sciences; our own motivation is  the qualitative analysis of the fate of \emph{ecosystems}, see~\cite{pommereau2022}.

Multistability can be succinctly described as the presence of several \emph{attractors} in the system under study.
Attractors characterize the stable behaviours, given as the smallest subsets of states from which the system cannot escape;  in other words, they are {terminal strongly connected components} of the associated transition system. In the long run, the system will enter one of its attractors and remain inside; multi-stability arises when there is more than one such attractor.

\textbf{Remark:}
Regardless of the application domain, there are typically some attractors that play a more `negative' role then others; in cell regulation, e.g., some attractors simply represent different healthy phenotypes into which a cell may differentiate, while others can be cancerous.
In ecology, there may even be not a single attractor that can be considered `healthy' in the sense that every such stable region may be characterized by the collapse of some species or sub-ecosystems. The survival, or the \emph{avoidance of doom} as we will call it, consists for such systems in staying forever in some transient but doom-free loop. The purpose of this article is to provide formal tools for addressing these forms or doom avoidance, in the context of concurrent systems modeled by safe Petri nets.

Returning to the discussion of basic notions, the \emph{basin} $\basin{\attn}$ of a given attractor $\attn$ consists of  the states that are \emph{doomed} in the sense that any infinite run from them inevitably leads the system  into $\attn$. 

The  basin includes the attractor itself, and possibly one or several transient states~\cite{Klarner18}.

We aim at finding the basin boundaries at which the system switches from an undetermined or \emph{free} state into some basin. While interesting beyond that domain, this is a recurrent question in the analysis of 
 signalling and gene regulatory networks~\cite{Cohen2015,Mendes2018}.
In \cite{bifurcations-BMC}, the authors provide a method for identifying, in a boolean network model, the states in which one transition
leads to losing the reachability of a given attractor (called \emph{bifurcation transitions} there; we prefer to speak of \emph{tipping points} instead).
However, enumerating the states in which the identified transitions make the system branch away from the attractor can be highly combinatorial and hinders a fine understanding of the branching.
Thus, the challenge resides in identifying the specific contexts and sequences of transitions leading to a strong basin.

\emph{Unfoldings} of Petri nets \cite{Esparza08}, which are essentially event structures in the sense of Winskel et al.~\cite{NPW80} with additional information about \emph{states},
are an acyclic representation of the possible sequences of transitions, akin to Mazurkiewicz traces 
but enriched with branching information.

Many reachability-related verification problems for concurrent systems have been successfully addressed by Petri-net unfolding methods over the past decades, see~\cite{McM92,ERV02,Esparza08}. 
However, questions of long-term behaviour and stabilization have received relatively little attention.

 We have shown in previous work~\cite{CHJPS-cmsb14} how all reachable  attractors can be extracted using  bounded unfolding prefixes.  Also, we have exhibited (\cite{HPS-cmsb20}) the particular shape of basins that can arise in a concurrent model.

In the present paper, we build on these previous results; 
the point of view taken here is that all attractors correspond to the \emph{end} of the system's free behaviour, in other words to its \emph{doom}. We will give characterizations of  basin boundaries (called \emph{cliff-edges} below), and  of those behaviours that remain \emph{free}, in terms of properties of the unfolding, reporting also on practical experiments with an implementation of the algorithms derived.  We finally introduce a novel type of quantitative measure, called \emph{protectedness}, 
to indicate how far away (or close) a system is from doom, in a state that is still free per se. General discussions and outlook will conclude this paper.

\section{Petri Nets and Unfoldings}
\label{sec:unf}
We begin now by recalling the basic definitions needed below.
A \textbf{Petri net} is a bipartite directed graph whose nodes are either \emph{places} or \emph{transitions}, and places may carry \emph{tokens}.
In this paper, we consider only \emph{safe} Petri nets where  a place carries either one or no token in any reachable marking. 
The set of currently active places form the state, or \emph{marking}, of the net.

\textbf{Note.} Some remarks are in order concerning our use of Petri nets versus that of  \emph{boolean networks}, which are more widely used in systems biology. 
Safe (or 1-bounded) Petri nets~\cite{Mur89} are close to Boolean and multivalued networks~\cite{CHKPT-nc19}, yet  enable a more fine-grained specification of the conditions for triggering value changes. Focussing on safe PNs entails no limitation of generality of the model, as two-way behaviour-preserving translations between Boolean and multivalued models exist (see~\cite{CHKPT-nc19} and the appendix of~\cite{CHJPS-cmsb14} for discussion). We are thus entitled to move between these models without loss of expressiveness; however, Petri nets provide more convenient ways to develop and present the theory and the algorithms here.

Formally, a \emph{net} is a tuple $\net=\tup{\place,\trans,\flow}$,  where  $\trans$ is a finite set of  \emph{transitions}, $\place$ a finite set of \emph{places}, 
  and  $\flow\subseteq(\place\times \trans)\cup(\trans\times \place)$ is a \emph{flow relation} whose elements are called \emph{arcs}. 
  In figures,  places are represented by circles and the transitions by boxes (each one with a label identifying it).

For any node $x\in \place\cup \trans$, we call \emph{pre-set} of $x$ the set $\preset{x}=\{y\in \place\cup \trans\mid \tup{y,x}\in \flow\}$ and \emph{post-set} of $x$ the set $\postset{x}=\{y\in \place\cup \trans\mid \tup{x,y}\in \flow\}$.
A \emph{marking} for $\net$ is a mapping $\markn:\place\to\NN_0$.
A \emph{Petri net} is a tuple $\petrinet=\tup{\place,\trans,\flow,\markn_0}$, with $\markn_0\subseteq\place$ {(i.e., $\markn_0:\place\to\{0,1\} $)} an \emph{initial marking}.
Markings are represented by dots (or tokens) in the marked places.
A transition $\transn\in \trans$ is \emph{enabled} at a marking $\markn$, denoted $\markn\move{\transn}$, if and only if $\forall \placen\in\preset{\transn}:~\markn(\placen)\geq 1$.
An enabled transition $\transn$ can \emph{fire}, leading to the new marking $\markn'$ given by $\markn'(\placen)=(\markn(\placen)-\one_{\preset{\transn}}(\placen))+ \one_{\postset{\transn}}(\placen)$
in that case we write $\markn\move{\transn}\markn'$.
A \emph{firing sequence} from a marking $M'_0$ is a (finite or infinite) sequence $w=t_1t_2t_3\dots$ over $\trans$ such that there exist markings $\markn'_1,\markn'_2,\dots$ with $\markn'_0\move{t_1}\markn'_1\move{t_2}\markn'_2\move{t_3}\dots$. 
If $w$ is finite and of length $n$, we write $\markn'_0\move{w}\markn'_n$, and we say that $\markn'_n$ is \emph{reachable} from $\markn'_0$, also simply written $\markn'_0\rightarrow\markn'_n$. We denote the set of markings reachable from some marking $\markn$ in a net $\net$ by $\reach_\net(\markn)$.

A \emph{cycle}  is a tuple $\tup{\markn_1,\ldots,\markn_{n}}$ of reachable markings such that there exists a finite firing sequence $w=t_1t_2\ldots t_{n}$ for which $\markn_1\move{t_1}\markn_2\move{t_2}\markn_3\move{t_3}\ldots\move{t_{n}}\markn_1$. Clearly, $\tup{\markn_1,\ldots,\markn_{n}}$  is a cycle  iff $\tup{\markn_2,\ldots,\markn_{n},\markn_1}$ is.
A Petri net $\tup{\net,\markn_0}$ is \emph{$n$-bounded} if $\markn(\placen)\leqslant n$ for  every reachable marking $\markn\in\reach_\net(\markn_0)$ and every place $\placen\in\place$. A $1$-bounded Petri net is called \emph{safe}.
In this paper, we assume that all  Petri nets  considered are safe. 

From an initial marking of the net, one can recursively derive all possible
transitions and reachable markings, resulting in a \emph{marking graph}
(Def.~\ref{def:marking-graph}).

\begin{definition}
\label{def:marking-graph}
Let $\net=\tup{\place,\trans,\flow}$ be a net and $\states$ a set of markings. The \emph{marking graph} induced by $\states$ is
a directed graph $\tup{\states,\edges}$ such that $\edges\subseteq \states\times \states$ contains $\tup{\markn,\markn'}$ iff $\markn\move{t}\markn'$ for some $\transn\in \trans$; 
the arc $\tup{\markn,\markn'}$ is then \emph{labeled} by $\transn$.
The \emph{reachability graph} of a Petri net $\tup{\net,M_0}$ is the graph induced by $\reach_\net(\markn_0)$.
\end{definition}

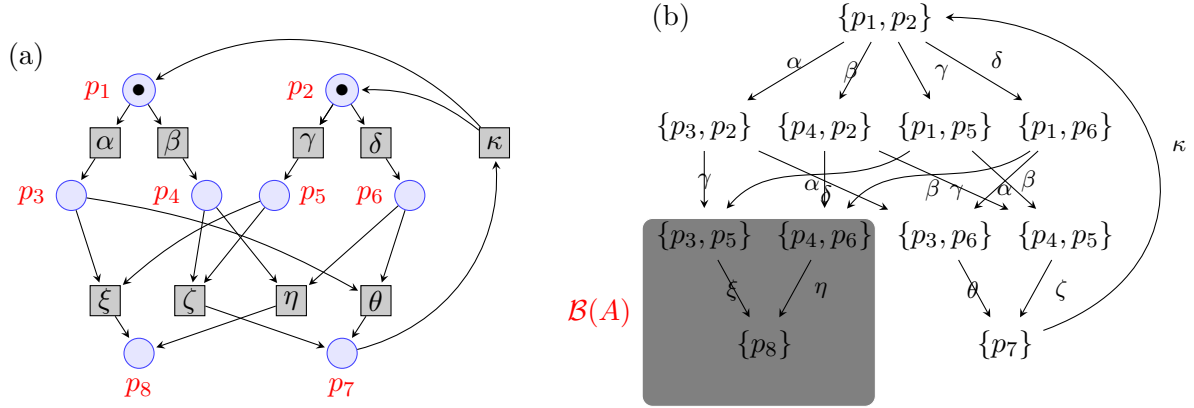
\begin{figure}[ht]
    \centering
    \begin{subfigure}[t]{0.48\textwidth}
    \def\a{1}
\def\b{0.45}
\def\c{0.7}
\begin{tikzpicture}[>=stealth,shorten >=1pt,node distance=\a cm,auto]
  \node[cond] (p1) at (0, 0) [label=left:\text{$p_1$}] {$\bullet$};
  \node[cond] (p2) at (6*\b, 0)  [label=left:\text{$p_2$}] {$\bullet$};
  \node[cond] (p3) at (-2*\b, -2*\c)  [label=left:\text{$p_3$}] {};
  \node[cond] (p4) at (2*\b, -2*\c)  [label=left:\text{$p_4$}] {};
  \node[cond] (p5) at (4*\b, -2*\c)  [label=right:\text{$p_5$}] {};
  \node[cond] (p6) at (8*\b, -2*\c)  [label=left:\text{$p_6$}] {};
  \node[cond] (p7) at (6*\b, -5*\c)  [label=below:\text{$p_7$}] {};
  \node[cond] (p8) at (0*\b, -5*\c)  [label=below:\text{$p_8$}] {};
  
  \node[transition] (a) at (-1*\b, -\c) {$\alpha$};
  \node[transition] (b) at (1*\b, -\c) {$\beta$};

  \node[transition] (x) at (-1*\b, -4*\c) {$\xi$};
\node[transition] (c) at (5*\b, -\c) {$\gamma$};
\node[transition] (d) at (7*\b, -\c) {$\delta$};
\node[transition] (e) at (7*\b, -4*\c) {$\theta$};
\node[transition] (r) at (10.5*\b, -1*\c) {$\kappa$};
\node[transition] (f) at (4.5*\b, -4*\c) {$\eta$};
\node[transition] (g) at (1.5*\b, -4*\c) {$\zeta$};
 \path[->] (p1)  edge (a)
        (p3)  edge (x)
        (p3)  edge [bend left=10]    (e)
         (p1)  edge (b)
         (p2) edge (c)
         (p2) edge (d)
         (c) edge (p5)
        (p5) edge (g)
        (p5) edge [bend right=10] (x)
         (p4) edge (g)
         (p4) edge (f)
        (p6) edge (f)
          ; 
 \path[->] (p7) edge [bend right=40] (r)
           (r) edge [bend right=20] (p2) 
           (r) edge [bend right=40] (p1)
 ;
  \path[->]  (a) edge (p3)
  (x) edge (p8)
   (b) edge (p4)
    (d) edge (p6)
    (p2) edge (c)
(p6) edge  (e)
 (e) edge (p7)
 (f) edge (p8)
  (g) edge (p7)  
 ;
 \node at (-1.5*\a,0.45*\a) {(a)\phantomsubcaption\label{fig:basinexample}};
\end{tikzpicture}
    \end{subfigure}
    \begin{subfigure}[t]{0.48\textwidth}
    \def\xdis{1.6cm}
\def\ydis{1.45cm}
\begin{tikzpicture}[>=stealth,shorten >=1pt,node distance=\ydis and \xdis,on grid]
	\node (12) {$ \{p_1, p_2 \} $};
	\node[below =of 12,xshift=-0.5*\xdis] (42) {$ \{p_4, p_2 \} $};
	\node[left=of 42] (32) {$ \{p_3, p_2 \} $};
	\node[below=of 12,xshift=0.5*\xdis] (15) {$ \{p_1, p_5 \} $};
	\node[right=of 15] (16) {$ \{p_1, p_6 \} $};
	\node[below=of 32] (35) {$ \{p_3, p_5 \} $};
	\node[below=of 42] (46) {$ \{p_4, p_6 \} $};
	\node[below=of 15] (36) {$ \{p_3, p_6 \} $};
	\node[below=of 16] (45) {$ \{p_4, p_5 \} $};
	\node[below=of 46,xshift=-0.5*\xdis] (8) {$ \{p_8 \} $};
	\node[below=of 45,xshift=-0.5*\xdis] (7) {$ \{p_7 \} $};
 \node[none,  left of = 7](aux1){};
\ node[below left, inner sep= 0.5mm] {\footnotesize$\delta$}
 \begin{scope}[on background layer]
    \node[grayzone, left of  =aux1] (A1) [label = left:$A$] {};  
 \end{scope}
 \node[none,  left of = 36](aux2){};
 \begin{scope}[on background layer]
    \node[grayzone2, below left of  =aux2] (A1) [label = left:$\mathcal{B}(A)$] {};
 \end{scope}	
	\path[->]
	(12) 	edge node[above left, inner sep= 0.5mm] {\footnotesize$\alpha$} (32)
			edge node[left, inner sep= 1mm] {\footnotesize$\beta$} (42)
			edge node[right, inner sep= 1mm] {\footnotesize$\gamma$} (15)
			edge node[above right, inner sep= 0.5mm] {\footnotesize$\delta$} (16)
	(32)	edge node[left, inner sep= 0.5mm] {\footnotesize$\gamma$}
 (35)
		edge 
   (36)
	(42)	edge node[below, inner sep= 0.5mm] {\footnotesize$\gamma$} (45)
			edge
   node[below left, inner sep= 0.5mm] {\footnotesize$\delta$}
   (46)
	(15)	edge[out=-145,in=50]  node[below left, inner sep= 0.5mm] {\footnotesize$\alpha$} (35)
			edge node[right, inner sep= 0.5mm] {\footnotesize$\beta$}(45)
	(16)	edge node[below
 left,  inner sep= 0.5mm] {\footnotesize$\alpha$} (36)
			edge[out=-145,in=50] node[below left, inner sep= 0.5mm] {\footnotesize$\beta$} (46)
	(35)	edge node[left, inner sep= 1mm] {\footnotesize$\xi$} (8)
	(46)	edge node[right, inner sep= 1mm] {\footnotesize$\eta$} (8)
	(36)	edge node[left, inner sep= 1mm] {\footnotesize$\theta$}  (7)
	(45)	edge node[right, inner sep= 1mm] {\footnotesize$\zeta$} (7)
	(7)		edge[out=20, in=0, looseness=1.7] node[right, inner sep= 1mm] {\footnotesize$\kappa$} (12)
	;
 \node at (-3,0) {(b)\phantomsubcaption\label{fig:basinstategr}};
\end{tikzpicture}
    \end{subfigure}
    \captionsetup{subrefformat=parens}
    \caption{Petri net example from~\cite{HPS-cmsb20} in \subref{fig:basinexample}, and its reachability graph in \subref{fig:basinstategr}.  The only attractor $A$ is highlighted in dark gray, and its (strong) basin $\mathcal{B}(A)$ in light gray.
    \label{fig:teamplay}
    }
\end{figure}

Figure~\ref{fig:basinstategr} shows the reachability graph for our running example \ref{fig:basinexample}.
Note that the reachability graph is always finite for safe Petri nets.

\textbf{Unfoldings.}
Roughly speaking, the unfolding of
a Petri net $\petrinet$ is an acyclic Petri net (with particular structural properties), $\unfolding$ exhibiting  reproduces exactly the
same non-sequential  behaviours as $\petrinet$. 

Let us now give the technical definitions to introduce unfoldings formally. A  more extensive treatment  can be found, e.g., in \cite{ERV02,Esparza08}.

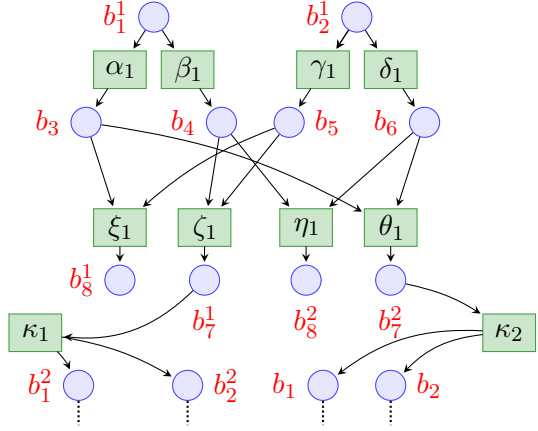
\begin{figure}[ht]
    \centering
\def\a{1}
\def\b{0.45}
\def\c{0.7}
\begin{tikzpicture}[>=stealth,shorten >=1pt,node distance=\a cm,auto]
  \node[cond] (b1) at (0, 0)  [label=left:\text{$b_1^1$}] {};
  \node[cond] (b2) at (6*\b, 0)  [label=left:\text{$b_2^1$}] {};
  \node[cond] (b1') at (5*\b, -7*\c)  [label=left:\text{$b_1$}] {};
  \node[cond] (b2') at (7*\b, -7*\c)  [label=right:\text{$b_2$}] {};
  \node[cond] (b1'') at (-1, -7*\c)  [label=left:\text{$b_1^2$}] {};
  \node[cond] (b2'') at (1*\b, -7*\c)  [label=right:\text{$b_2^2$}] {};
  \node[cond] (b3) at (-2*\b, -2*\c)  [label=left:\text{$b_3$}] {};
  \node[cond] (b4) at (2*\b, -2*\c)  [label=left:\text{$b_4$}] {};
  \node[cond] (b5) at (4*\b, -2*\c)  [label=right:\text{$b_5$}] {};
  \node[cond] (b6) at (8*\b, -2*\c)  [label=left:\text{$b_6$}] {};
  \node[cond] (b7a) at (1.5*\b, -5*\c)  [label=below:\text{$b_7^1$}] {};
  \node[cond] (b7b) at (7*\b, -5*\c)  [label=below:\text{$b_7^2$}] {};
  \node[cond] (b8a) at (-1*\b, -5*\c)  [label=left:\text{$b_8^1$}] {};
 \node[cond] (b8b) at (4.5*\b, -5*\c)  [label=below:\text{$b_8^2$}] {};

  \node[event] (a) at (-1*\b, -\c) {$\alpha_1$};
  \node[event] (b) at (1*\b, -\c) {$\beta_1$};

  \node[event] (x) at (-1*\b, -4*\c) {$\xi_1$};
\node[event] (c) at (5*\b, -\c) {$\gamma_1$};
\node[event] (d) at (7*\b, -\c) {$\delta_1$};
\node[event] (e) at (7*\b, -4*\c) {$\theta_1$};
\node[event] (r) at (10.5*\b, -6*\c) {$\kappa_2$};
\node[event] (rbis) at (-3.5*\b, -6*\c) {$\kappa_1$};
\node[event] (g) at (1.5*\b, -4*\c) {$\zeta_1$};
\node[event] (f) at (4.5*\b, -4*\c) {$\eta_1$};
\node (phantom1) at (-1, -8*\c)  [] {};
\node (phantom2) at (1*\b, -8*\c)  [] {};
\node (phantom3) at (5*\b, -8*\c)  [] {};
\node (phantom4) at (7*\b, -8*\c)  [] {};
 \path[->] (b1)  edge (a)
        (b3)  edge (x)
        (b3)  edge [bend left=10]    (e)
         (b1)  edge (b)
         (b2) edge (c)
         (b2) edge (d)
         (c) edge (b5)
        (b5) edge (g)
        (b5) edge [bend right=10] (x)
         (b4) edge (g)
         (b4) edge (f)
        (b6) edge (f)
          ; 
 \path[->] (b7b) edge [bend left=10] (r)
           (r) edge [bend right=20] (b2') 
           (r) edge [bend right=20] (b1')
 (b7a) edge [bend left=25] (rbis)
           (rbis) edge [bend left=10] (b2'') 
           (rbis) edge [bend left=10] (b1'')
 ;
  \path[->]  (a) edge (b3)
  (x) edge (b8a)
   (b) edge (b4)
    (d) edge (b6)
    (b2) edge (c)
(b6) edge  (e)
 (e) edge (b7b)
 (f) edge (b8b)
  (g) edge (b7a)  
 ;
 \draw[densely dotted, thick] (b1') -- (phantom3);
 \draw[densely dotted, thick] (b2') -- (phantom4);
 \draw[densely dotted, thick] (b1'') -- (phantom1);
 \draw[densely dotted, thick] (b2'') -- (phantom2);
\end{tikzpicture}
\caption{A prefix of the unfolding for the Petri net of Figure~\ref{fig:basinexample}.
    \label{fig:basinunfold}}
\end{figure}

\begin{definition}[Causality, conflict, concurrency]
\label{def:causality}
Let $\net=\tup{P,T,F}$ be a net and $x,y\in P\cup T$ two nodes
of $\net$.
We say that
$x$ is a \emph{causal predecessor} of~$y$, noted $x<y$, if there exists a
non-empty path of arcs from $x$ to $y$. We note $x\leq y$ if $x<y$ or $x=y$.
If $x\leq y$ or $y\leq x$, then $x$ and $y$ are said to be
\emph{causally related}.
Transitions $u$ and $v$ are in \emph{direct conflict}, noted $u\dircf v$, iff $\preset{u}\cap\preset{v}\ne
\emptyset$; nodes
$x$ and $y$ are \emph{in conflict}, noted $x\conflict y$, if there exist
$u,v\in T$ such that $u\ne v$, $u\le x$, $v\le y$, and $u\dircf v$. We call
$x$ and $y$ \emph{concurrent}, noted $x\concurrent y$, if they are neither
causally related nor in conflict.
A set of concurrent places is called a \emph{co-set}.
\end{definition}
In Figure~\ref{fig:basinunfold}, $\alpha_1$ and $\beta_1$ are in conflict,  while $\alpha_1$ and $\gamma_1$ are concurrent. Further, $\alpha_1$ is a causal predecessor of $\xi_1$, $\theta_1$, and $\kappa_2$; readers will easily identify other relations in this figure.
\begin{definition}[Occurrence net]
\label{def:occurrencenet}{\it
Let $\onet=\tup{\oplace,\otrans,\oflow,\cutn_0}$ be a Petri net. We say that $\onet$
is an \emph{occurrence net} if it satisfies the following properties:
\begin{enumerate}
\item The causality relation $<$ is acyclic and well-founded;
\item $|\preset{\oplacen}|\le 1$ for all places $\oplacen\in\oplace$,
  and $\oplacen\in \cutn_0$ iff $|\preset{\oplacen}|=0$;
\item For every transition $\otransn\in\otrans$, $\otransn\conflict\otransn$ does not hold, and \(\{x \mid x \leq \otransn\}\) is finite.
\end{enumerate}}
\end{definition}

The reader is invited to check that the net in Figure~\ref{fig:basinunfold} is indeed an occurrence net.

Following the  convention in the unfolding literature, we refer to the
places of an occurrence net as \emph{conditions} ($\oplace$) and to its transitions
as \emph{events} ($\otrans$). 
Due to the structural constraints, the firing
sequences of occurrence nets have special properties: if some condition $\oplacen$ is
marked during a run, then the token on $\oplacen$ was either present initially in $\cutn_0$,
or produced by one particular event (the single event in~$\preset{\oplacen}$);
moreover, once the token on $\oplacen$ is consumed, it can never be replaced by
another token, due to acyclicity of~$<$. 
\begin{definition}[Configurations, cuts]
\label{def:configuration}{\it
Let $\onet=\tup{\oplace,\otrans,\oflow,\cutn_0}$ be an occurrence net.
A set $\confign\subseteq \otrans$ is called a
\emph{configuration} of $\onet$ if (i) $\confign$ is \emph{causally
closed}, i.e.\ $e'<e$ and $e\in \confign$ imply $e'\in \confign$; and (ii) $\confign$ is \emph{conflict-free}, i.e.\ if $e,e'\in\confign$,
then $\neg(e\conflict e')$. 
In particular, for any $e\in E$, $\cone{e}\define\{e'\in E:~e'\myleq e\}$ and $\trunk{e}\define\{e'\in E:~e'< e\}$ are configurations, called the \emph{cone} and \emph{stump} of $\otransn$, respectively; any $\confign$ such that $\exists~e\in E:~\confign=\cone{e}$ is called a \emph{prime configuration}.
Denote the set of all configurations of $\onet$ as $\configs(\onet)$, and its subsets containing all \textbf{finite} configurations as $\finfigs(\onet)$, where we drop the reference to $\onet$ if no confusion can arise.  The \emph{cut} of a finite $\confign$, denoted $\cut(\confign)$, is the set of
conditions $(\cutn_0\cup\postset{\confign})\setminus\preset{\confign}$. 
A \emph{run} is a maximal element of $\configs(\onet)$ w.r.t.\ set inclusion; denote the set of $\onet$'s runs as $\runs=\runs(\onet)$, and its elements generically by $\runn$.
 Denote by ${\hinfigs}(\onet)$ the set of all \emph{infinite} configurations, and let 
 \begin{equation*}
     \infigs(\onet)\define{\hinfigs}(\onet) \cup\runs\left(\onet\right)
 \end{equation*}
If $\confign\in\finfigs$, let the \emph{crest} of $\confign$ be the set 
$\crest{\confign}\define \max_<(\confign)$ of its maximal events. 
We say that configuration $\confign$ \emph{enables} event $\otransn$, written $ \confign\omove{\otransn}$, iff i) $\otransn\not\in\confign$ and ii) $\confign\cup\{\otransn\}$ is a configuration.
} 
Configurations $\confign_1,\confign_2$ are in conflict, written $\confign_1\conflict\confign_2$, iff $(\confign_1\cup\confign_2)\not\in\configs$ or, equivalently, iff there exist $\otransn_1\in\confign_1$ and $\otransn_2\in\confign_2$ such that $\otransn_1\conflict\otransn_2$.\footnote{The use of the same symbol $\conflict$ is motivated by the fact that $\confign_1=\cone{\otransn_1}$ and $\confign_2=\cone{\otransn_2}$ implies $\confign_1\conflict\confign_2\Leftrightarrow \otransn_1\conflict\otransn_2$.}
Let  $\confign$ be  a configuration and  $E\cap\confign=\emptyset$ such that $\confign\cup E$ is a configuration.  Write $\confign\oplus E\define \confign\cup E$ in that case; we call $\confign\oplus E\define \confign\cup E$ an \emph{extension} of $\confign$, and $E$ a \emph{suffix} of $\confign$.
\end{definition}

Intuitively, a  configuration is the partially ordered set of transition firings  occurring during
an enabled firing sequence of $\petrinet$, and its cut (if it exists) is the set of conditions
marked after completing that firing sequence. Note that $\emptyset$ is a configuration, that $\crest{\emptyset}=\emptyset$,
and that $\cutn_0$ is the cut of the configuration $\emptyset$.
Moreover, if $\infigs(\onet)\ne\emptyset$, then  $\infigs(\onet)\cap\runs(\onet)\ne\emptyset$; however, it is in general not the case that $\infigs(\onet)\subseteq\runs(\onet)$. 
The crest of a prime configuration $\cone{e}$ is $\{e\}$.

In Figure~\ref{fig:basinunfold}, the initial cut is $\cutn_0=\{b_1^1,b_2^1\}$;
we have prime configurations, e.g., $\{\alpha_1\}$, $\{\beta_1\}$, $\{\alpha_1,\gamma_1,\xi_1\}$, $\{\beta_1,\gamma_1,\zeta_1\}$ etc, 
and non-prime configurations $\{\alpha_1,\gamma_1\}$, $\{\alpha_1,\delta_1\}$ etc.

\textbf{Definition of Unfoldings.}

\begin{definition}[Net homomorphism] Let $\net_1=\tup{\place_1,\trans_1,\flow_1}$ and $\net_2=\tup{\place_2,\trans_2,\flow_2}$ be two nets. A \emph{homomorphism} from $\net_1$ to $\net_2$ is a mapping $\morph:\place_1\cup\trans_1\to\place_2\cup\trans_2$ such that $\morph(\place_1)\subseteq \place_2$ and $\morph(\trans_1)\subseteq\trans_2$, and that satisfies, in addition, for every $\transn\in\trans_1$,
\begin{eqnarray*}
\morph\left(\preset{\transn}\right)=\preset{\morph(\transn)}&\mathit{and}&\morph\left(\postset{\transn}\right)=\postset{\morph(\transn)}
\end{eqnarray*}
\end{definition}
\begin{definition}
    [Branching Process and Unfolding] Let $\petrinet=\tup{\place,\trans,\flow,\markn_0}$
be a safe
Petri net. A \emph{branching process} of $\petrinet$ is a pair $\Pi=(\onet,\fold)$ with $\onet=\tup{\oplace,\otrans,\oflow,\cutn_0}$ an occurrence net and $\fold:\oplace\cup\otrans\to\place \cup \trans$ a homomorphism that satisfies the following parsimony property:
\begin{equation}
    \label{eq:parsi}
    \forall~\otransn,\otransn'\in\otrans:~
    \left.\begin{array}{rcl}
       \preset{\otransn}  &=&\preset{\otransn'}  \\
       \fold(\otransn)&=  & \fold(\otransn')
    \end{array}
    \right\} \Rightarrow \otransn =\otransn'
\end{equation}
If $\Pi_1=(\onet_1,\fold_1)$ and $\Pi_2=(\onet_2,\fold_2)$ are two branching processes of $\petrinet$, we say that $\Pi_1$ is a \emph{prefix} of $\Pi_2$ iff i) $\onet_1$ is a prefix modulo net isomorphism of $\onet_2$, and ii) $\fold_1$ agrees on its domain, modulo net isomorphism, with $\fold_2$'s restriction to $\fold_1'$s domain. There exists  a unique (up to isomorphism)  branching process $\Pi_\unfolding=(\unfolding,\fold_\unfolding)$ that is maximal in the sense that any branching process of which $\Pi_\unfolding$ is a prefix, must be an isomorphic copy of $\Pi_\unfolding$; we call \emph{unfolding of $\petrinet$} the maximal branching process $\Pi_\unfolding$ and, by
abuse of terminology, also the occurrence net $\unfolding$ if no confusion can arise.
\end{definition}

In the unfolding $(\unfolding,\fold)$ of
$\petrinet$ with  $\unfolding=\tup{\oplace,\otrans,\oflow,\cutn_0}$,
the firing sequences and reachable cuts of $\unfolding$ correspond exactly
the firing sequences and reachable markings of $\petrinet$ under the homomorphism $\fold$, see below.
Note that the occurrence net $\unfolding$ may be infinite; it can be inductively constructed as follows:
\begin{enumerate}
\item Every condition in $\oplace$ is characterized by a pair $(\otransn,\placen)\in (\otrans\cup\{\bot\})\times \place$.
 For  condition $\oplacen=\tup{\otransn,\placen}$, we will have $\otransn=\bot$ iff $\oplacen\in \cutn_0$;
 otherwise $\otransn$ is the singleton event in $\preset{\oplacen}$. Moreover, $\fold(\oplacen)=\placen$.
 The initial cut $\cutn_0$ contains as many  conditions $\tup{\bot,\placen}$
 for each token initially on place $\placen$ under $\markn_0$ in $\petrinet$.
\item The events of $\otrans$ are a subset of $2^\oplace\times \trans$. 
More precisely, for every co-set $\oplace'\subseteq\oplace$ such that
$\fold(\oplace')=
\preset{\transn}$,
we have an event $\otransn=\tup{\oplace',t}$.
 In this case, we add edges $\tup{\oplacen,\otransn}$
 for each $\oplacen\in \oplace'$ (i.e. $\preset{\otransn}=\oplace'$), we set $\fold(\otransn)=\transn$, and for each
 $\placen\in\postset{\transn}$, we add to $\oplace$ a condition $\oplacen=\tup{\otransn,\placen}$ connected by
 an edge $\tup{\otransn,\oplacen}$.
\end{enumerate}
Intuitively, a condition $\tup{\otransn,\placen}$ represents the possibility of putting
a token onto place $\placen$ through a particular set of events, while an event
$\tup{\oplace',\transn}$ represents a possibility of firing transition $\transn$ in a particular
context.

\textbf{Configurations and Markings.}
The following facts from the literature  will be useful:\begin{Lemma}[see e.g.~\cite{ERV02}]
\label{le:safe}{\it Fix $\petrinet=\tup{\place,\trans,\flow,\markn_0}$  and its unfolding $\unfolding=\tup{\oplace,\otrans,\oflow,\cutn_0, \fold}$.
 Then for any two conditions (events) $\oplacen,\oplacen'$ ($\otransn,\otransn'$) such that $\oplacen\concurrent\oplacen'$ ($\otransn \concurrent \otransn'$), one has $\fold(\oplacen)\ne\fold(\oplacen')$ ($\fold(\otransn)\ne\fold(\otransn')$). 
Moreover, every
 finite configuration $\confign$ of $\unfolding$ represents a possible firing
sequence whose resulting marking corresponds, due to the construction of
$\unfolding$, to a reachable marking of $\petrinet$. This marking is defined
by
$\marking{\confign}\define \fold^{-1}(\cut(\confign))$.
Moreover, for any two distinct configurations $\confign_1,\confign_2$ that satisfy $\marking{\confign_1}=\marking{\confign_2}$, we have an isomorphism of labeled occurrence nets
\begin{equation}\label{eq:restart}
    \isom_{(\confign_1,\confign_2)}:\unfolding_{\slash \confign_1}\to \unfolding_{\slash \confign_2},
\end{equation}
where 
$\unfolding_{\slash \confign_1}$ is the suffix of $\unfolding$ after removing configuration $\confign_1$ and all nodes in conflict with $\confign_1$.}
\end{Lemma}
In fact, $\unfolding_{\slash \confign_1}$ and $\unfolding_{\slash \confign_2}$ in (\ref{eq:restart}) are isomorphic copies of 
$\unfolding (\netn,\marking{\confign_1})$.
This means, informally speaking, that any configuration of the system can be split into consecutive parts in such a way that each part is itself a configuration obtained by unfolding the Petri net `renewed' with the marking reached by the previous configuration.

\textbf{Complete Prefix.}
In general, $\unfolding$
is an infinite net, but if $\petrinet$ is bounded, then it is possible
to compute a finite prefix $\prefn$
of $\unfolding$ that is ``complete''
in the sense that every reachable marking of $\petrinet$ has a reachable
counterpart in $\prefn$,  and vice 
versa. One may require other completeness properties, as we will see below; here, the definition follows the notion dominant in the literature.
\begin{definition}[complete prefix, see \cite{McM92,ERV02,Esparza08}]
\label{def:complete}
Let $\petrinet=\tup{\net,\markn_0}$ be a bounded Petri net and
$\unfolding=\tup{\oplace,\otrans,\oflow,\cutn_0,\fold}$
its unfolding. A finite occurrence net ${\prefn}=\tup{\oplace',\otrans',\oflow',\cutn_0}$
is said to be a \emph{prefix} of $\unfolding$ if $\otrans'\subseteq \otrans$ is
causally closed, $\oplace'=\cutn_0\cup\postset{\otrans'}$, and $\oflow'$ is the restriction
of $\oflow$ to $\oplace'$ and $\otrans'$. A prefix $\prefn$ is said to be \emph{complete}
if for every reachable marking $\markn$ of $\petrinet$ there exists a
configuration $\confign$ of $\prefn$ such that (i) $\marking{\confign}=\markn$,
and (ii) for each transition $\transn\in \trans$ enabled in $\markn$, there is an event
$\tup{\oplace'',\transn}\in \otrans'$ enabled in $\cut(\confign)$. 
\end{definition}

We shall write $\prefixn_0=\prefixn_0(\petrinet)$ to denote an arbitrary complete prefix of the unfolding of $\netn$.
The construction of such a complete prefix
is indeed possible (\cite{McM92,ERV02}), and efficient tools such as \textsc{Ecofolder} (\cite{ecofolder2024}) and \textsc{Mole} (\cite{mole}) exist for this purpose.
Several ingredients of this construction  will play a role below, so we sketch them here.

\textbf{Cutoff events and the complete prefix scheme.} The unfolding is stopped on each branch when some \emph{cutoff event} is added.

The criterion for classifying an event $\otransn$ as cutoff is given by  marking equivalence: the marking $\marking{\cone{\otransn}}$ that $\otransn$ `discovers' has already been discovered by a \emph{smaller} (wrt some ordering relation $\adequate$) configuration. Now, to ensure completeness, the ordering relation $\adequate$ to compare two configurations must be an \emph{adequate} order, in the following sense:
\begin{definition}[\cite{ERV02}, Def. 4.5]
A partial order $\adequate$ on $\finfigs$ is called \emph{adequate order} iff
\begin{itemize}
    \item $\adequate$ is well-founded, 
    \item $\confign_1\subseteq\confign_2$ implies $\confign_1\adequate\confign_2$, and \item $\adequate$ preserves extensions, i.e. for any $\confign_1\adequate\confign_2$ such that $\marking{\confign_1}=\marking{\confign_2}$, one has $\confign_1\oplus E\adequate \confign_2\oplus\isom_{(\confign_1,\confign_2)}(E)$ for the isomorphism $\isom_{(\confign_1,\confign_2)}$ from (\ref{eq:restart}).
\end{itemize}
\end{definition}   As shown in~\cite{ERV02}, for some  choices of $\prec$, the obtained prefix may be bigger than the reachability graph for some safe nets; however, if $\prec$ is a \emph{total} adequate order, the number of non-cutoff events of the prefix $\prefixn_0$ thus obtained never exceeds the size of the reachability graph.

We will refer  throughout this paper to the  complete  prefixes $\Pi_0^\adequate$  computed according to some adequate \emph{total} order, as is done in the tools \textsc{Mole}~\cite{mole} and \textsc{Ecofolder}~\cite{ecofolder2024}, as \emph{Esparza prefixes}. If instead one chooses $\adequate=\subsetneq$, the resulting prefix $\Pi^{McM}\define\Pi^\subsetneq_0$, as \textit{can} be done in \textsc{Ecofolder}~\cite{ecofolder2024}, is referred to as the \emph{McMillan prefix}.

\section{Attractors, Basins,  And Fairness}
\subsection{Attractors and Basins}\begin{definition}
\label{def:attr}
An \emph{attractor} $\attn\subseteq 2^\places$ is a terminal SCC of the marking graph; that is,
 two states in $\attn$ are reachable from one another, and no state outside $\attn$ is reachable from any state in $\attn$. Denote the set of attractors of $\netn$ reachable from a marking $\markn$ in $\net$ by $\atts$.
Attractor $\attn$ is   a \emph{fixed point} iff there is $\markn\in\states$ such that $\attn=\{\markn\}$,  and for any $\transn\in\trans$,  $\markn\move{\transn}\markn'$ implies $\markn=\markn'$.
\end{definition}

Unfoldings do not show attractors directly; however, the following observations are useful (fixing $\petrinet=(\netn,\markn_0)$ and $\attn\in\atts(\petrinet)$):
\begin{itemize}
    \item If some finite configuration $\confign$ satisfies $\marking{\confign}\in\attn$, then so does any finite configuration $\confign'$ such that $\confign\subseteq \confign'$. Write $\finfigs_\attn\define\{\confign\in\finfigs:~\marking{\confign}\in\attn\}$.
    \item In the light of the above, call a maximal run $\runn$ an \emph{$\attn$-run} iff there exists $\confign\in\finfigs_\attn$ such that $\confign\subseteq\runn.$ Denote the set of $\attn$-runs by $\runs_\attn$.
\end{itemize}
    \begin{definition}
    The $\configs$\emph{-basin} $\cbasin{\attn}$ of $\attn$ is the set of finite configurations all of whose maximal extensions land in $\attn$: 
    $\cbasin{\attn}\define\{\confign\in\finfigs:~\forall~\runn\in\runs:~\confign\subseteq\runn\Rightarrow\runn\in\runs_\attn\}$
        The \emph{basin} $\basin{\attn}$ of $\attn$ is the set of markings from which reaching $\attn$ is inevitable:
        $\basin{\attn}\define\{\markn\in\reach(\markn_0):~\forall\confign\in\finfigs(\petrinet):~\marking{\confign}=\markn\Rightarrow \confign\in\cbasin{\attn}\}$  
    \end{definition}

By definition,  any attractor is  an \emph{absorbing} set of states (and so is any basin); once a system run enters some attractor (some basin), it will stay there forever. A different question is whether any infinite execution will eventually enter some attractor basin; in general, the answer is negative, since the system may exhibit transient loops in which it can forever remain active without ever entering any basin. 

One of our central tasks below is to identify whether or not a system state allows to loop in such away as to avoid a fatal attractor's basin, or \emph{doom} as we will say. To clarify this point, and to close a gap in the literature on attractors, we will next discuss  which \emph{fairness} properties  prevent transient loops. Thus, the conclusion will be - as our title suggests - that, roughly speaking, fairness in behaviour may lead the system into doom. Put otherwise, avoiding doom requires to impose some sort of control in the system to prevent its free action from fatality.
\subsection{Fairness, and how it leads into an attractor basin}

\subsubsection*{Situation Fairness.}
Despite their name, attractors do not in any way `attract' the system's behaviour in their direction, nor is the system necessarily entering any attractor eventually. Standard examples are non-attractor loops in the state graph; restrictions to behaviour are needed to ensure that the system eventually leaves  such loops. Such an intuition is often  captured by the notion of \emph{(strong) fairness}, cf.~\cite{KindWalt,DBLP:journals/ipl/Vogler95}: any transition that is \emph{enabled} infinitely often, must also eventually occur.
It is often assumed that strong fairness is sufficient to guarantee that all maximal runs eventually enter one or another attractor (and will obviously stay in it). We report here that in concurrent systems, this assumption is false.   To see the point, consider
Figure~\ref{fig:concfair}. The Petri net's only attractor is formed by the marking $\{A\}$ (which coincides with its basin). However, the Petri net depicted might cycle forever in the set of states in which  $p_3$ and $p_4$  are never jointly marked, and therefore never enter  the attractor basin. 
 This shows that in order to ensure that the system eventually enters some attractor,  we need to restrict its behavior to those runs that `eventually explore all accessible branches'\footnote{Dually, of course, such behaviour is to be \emph{avoided} at any cost if it is undesirable to enter some attractor; we will return to this point in the next sections.}.

However,   strong fairness is not sufficient to ensure that a concurrent system eventually enters a terminal SCC;  the example of Figure~\ref{fig:concfair} illustrates this.

\begingroup
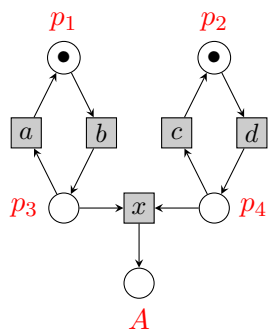
\begin{figure}[ht!]
  \centering
  \begin{tikzpicture}[>=stealth,node distance=1.0 cm]
    \node (p1) [state,label=90:$p_1$] {$\bullet$};
    \node (p2) [state,label=90:$p_2$,right of= p1,xshift=1.0cm] {$\bullet$};
    \node (p3) [state,label=180:$p_3$,below of= p1,yshift=-1.0cm] {};
    \node (p4) [state,label=0:$p_4$,below of= p2,yshift=-1.0cm] {};
    \node (p0) [state,label=270:$A$,right of= p1,yshift=-3cm] {};
    
    \node (t1) [transition, below of= p1, xshift=-.5cm] {\small $a$}
        edge [->]   (p1)
        edge [<-]   (p3);
    \node (t2) [transition, right of= t1] {\small $b$}
        edge [->]   (p3)
        edge [<-]   (p1);
    \node (t3) [transition, below of= p2, xshift=-.5cm] {\small $c$}
        edge [->]   (p2)
        edge [<-]   (p4);
    \node (t4) [transition, right of= t3] {\small $d$}
        edge [->]   (p4)
        edge [<-]   (p2);
    \node (t5) [transition, right of= p3] {\small $x$}
        edge [<-]   (p3)
        edge [<-]   (p4)
        edge [->]   (p0);
  \end{tikzpicture}
  \caption{Illustration of the fairness condition underlying the attractor notion. Clearly, the system has exactly one attractor, given by fixed point  $\attn\define\{\{A\}\}$. However, sequence $(badc)^\infty$ is strongly fair and never reaches $\attn$; on the other hand, every situation fair execution leads, after  finitely but unboundedly many steps, into $\attn$.}
  \label{fig:concfair}
\end{figure}
\endgroup
In this Petri net, a  finite firing sequence can be strongly fair only if its final marking is $\{A\}$; At the same time, no infinite strongly fair firing sequence in this net would permit $x$ to be enabled infinitely often. Nevertheless, there are infinite strongly fair executions that avoid enabling $x$ too often, e.g. the sequences  
$(badc)^\infty$ or  $(babadc)^\infty$, etc.
Such sequences, however, must necessarily be unfair to some transition \emph{in the context of the same marking}; in the example, the sequence $(badc)^\infty$ yields infinitely often the marking $\markn\define\{p_2,p_3\}$, and from this marking it `chooses' $a$ constantly, although $d$ is also enabled in $\markn$. Yet, strong fairness is fulfilled because $d$ fires infinitely often, but not from instances of $\markn$. 
To eliminate such `missed opportunities', we introduce a finer fairness notion. We will call an execution \emph{situation fair} iff any transition enabled in an infinitely visited marking, also fires infinitely often \emph{from that marking}. More formally:

\begin{definition}[Situation fairness] \label{def:sitfair}In $\netn$ as above, a firing sequence $\markn_0\move{\transn_1}\markn_1\move{\transn_2}\ldots$ is \emph{situation-fair} iff either (i) it is finite, and its last marking enables no transition, or (ii) for all $\transn\in \trans$ and all $\markn\subseteq \places$ such that $\markn\move{\transn}$: 
\begin{eqnarray}
\left|\left\{i\in \NN:~\markn_i=\markn\right\}\right|=\infty
&\Longrightarrow&
\left|\left\{j\in \NN:~\markn_j=\markn\land \transn_{j+1}={\transn}\right\}\right|=\infty.
\end{eqnarray}
\end{definition}
Note that such executions always exist; they may be obtained e.g. by applying a \emph{round robin} firing policy, in which, for $\{\transn_0,\ldots,\transn_{n-1}\}$ the transitions enabled at marking $\markn$, the transition  selected at the $k$-th visit to marking $\markn$ is  $\transn_{(k \mod n)}$. 
Before establishing the link between fairness and attractors, let us  introduce one more auxiliary notion:

\begin{definition}\label{de:distance}
For any reachable marking $\markn$, let $K_\markn$ be the smallest integer $k$ such that there exist an attractor $\attn$ and $\transn_1,\ldots,\transn_k\in\trans$ such that $\markn\move{\transn_1,\ldots,\transn_k}\markn_\attn$ with $\markn_\attn\in\attn$. Moreover, let $K_\netn\define\max_{\markn\in\reach(\markn)}(K_\markn)$.
\end{definition}
Note that in  safe Petri nets,
both $K_\markn$ and $K_\netn$ are well-defined and finite. 
Obviously $K_\markn$ must be finite for all reachable markings $\markn$ of $\netn$; 
since $\netn$ is finite and safe, $K_\netn$ is finite as well.

The following Theorem~\ref{le:sfattract} states that \emph{any} situation-fair execution of a safe net, i.e. round robin or other, will eventually leave any transient SCC and sooner or later enter a terminal SCC forever:
\begin{theorem}\label{le:sfattract}
Let $\sigma=\markn_0\move{\transn_1}\markn_1\move{\transn_2}\ldots$ be a situation-fair execution
of a safe Petri net $\netn$. Then either $\sigma$ is finite and its final marking is a fixed point, or $\sigma$ is infinite and there exists an attractor $\attn$ and $k\in\NN$ such that 
$\forall i\in\NN: \markn_{k+i}\in\attn$.
\end{theorem}
\bepr
By the definition of situation fairness, if $\sigma$ is finite, its final marking must be a fixed point. Thus assume that $\sigma$ is infinite; it then suffices to show that $\sigma$ eventually reaches an attractor, since by definition no markings outside the attractor are reachable from there. Since $\sigma$ is infinite and $\netn$ is safe, there must be at least one marking~$M$ that is visited infinitely often by $\sigma$. We shall prove that $M$ is part of an attractor. 
Indeed, suppose this is \emph{not} the case. Then $M$ is part of a transient SCC, and by definition, some attractor $\attn$ is reachable from $\markn$. Thus, we  have an executable path
$\markn\move{\transn_1}\markn_1\move{\transn_2}\ldots\move{\transn_n}\markn_n\in\attn$, for $n=K_M$ and some attractor $\attn$. Since $\sigma$ is situation fair vis-\`a-vis $\transn_1$ in $\markn_\sigma$, it must visit $\markn_1$ infinitely often as well. Repeating this argument, we obtain that all $\markn_k$, for $k=1,\ldots,n$, must be visited infinitely often, too. But $M$ is not reachable from $\markn_n$, which is a contradiction. Thus $M$ must be part of an attractor.
\eepr

We will now turn to the classification of states according to the long run behaviours available after them. That is, attractors may in general be desirable or undesirable; as long as the system still has some maximal behaviour available in which no \emph{bad} state is reached, we will call it \emph{free}, otherwise it is \emph{doomed}. The next section will make these notions precise.
\section{The Bad, the Good, the Doomed, and the Free 
}
\textbf{Bad states.} Our formal setting  contains and extends the one presented in~\cite{GiuaXie2005}, specialized here to the 1-safe case. We assume that we are given a set of \emph{bad markings} $\badstates\subseteq \reach(\markn_0)$, and write $\goodstates\define\reach(\markn_0)\backslash\badstates$. Since we are interested in long-term behaviours, we happily adopt the assumption from~\cite{GiuaXie2005} that $\badstates$ is reachability-closed, i.e. $\markn\in\badstates$ and $\markn\move{}\markn'$ imply $\markn'\in\badstates$.\newline
\textbf{Bad configurations.} Define $\badconfs\define\{\confign\in\finfigs:\marking{\confign}\in\badstates\}$ as the set of \emph{bad finite configurations}, and let $\badconfs^0$ be the set of configurations in $\badconfs$ that are contained in $\prefixn_0$. 
$\badconfs\subseteq\configs$  is \emph{absorbing} or \emph{upward closed}, that is,  
   for all $\confign_1\in\badconfs$ and $\confign_2\in\finfigs$ such that $\confign_1\subseteq\confign_2$,  one must have $\confign_2\in\badconfs$.

This upward closure justifies the following extension of our terminology:
let $\confign\in\infigs$; then $\confign$ is bad iff there exists 
$\confign'\in\badconfs$ such that $\confign'\subseteq\confign.$

For any  $\confign\in\configs$, let $\runs_\confign\define \left\{ \runn \in\runs:~\confign\subseteq \runn\right\}$ denote the maximal runs into which $\confign$ can evolve.
We are interested in 
those finite configurations all of whose extensions are `eventually bad'.

We will call such configurations \emph{doomed}, since from them, the system cannot avoid entering a bad marking sooner or later (and from then on, all reachable markings are bad).
\begin{definition}
\label{de:doom}{\it
Configuration $\confign\in\finfigs$ is
\begin{enumerate}
    \item 
\textbf{strongly
doomed}   iff
\begin{eqnarray}
\label{eq:latbad} \forall~\confign^*\in\infigs:~
\confign\subseteq
\confign^*~\Rightarrow~\exists~\confign'\in\badconfs:
\confign'\subseteq\confign^*
\end{eqnarray}
\item 
\textbf{doomed}   iff
\begin{eqnarray}
\label{eq:latbad2} \forall~\runn^*\in\runs:~
\confign\subseteq
\runn~\Rightarrow~\exists~\confign'\in\badconfs:
\confign'\subseteq\confign^*\label{eq:doom}
\end{eqnarray}
\end{enumerate}
Denote the set of strongly doomed configurations  by
$\doofigs^*$,  that of doomed configurations
by $\doofigs$, and the set of minimal elements in $\doofigs$ by $\mindofigs$.
We  call finite configurations that are not 
doomed 
\emph{free}. The set of 
free configurations is denoted by 
$\free\define\finfigs\backslash\doofigs$.}
\end{definition}
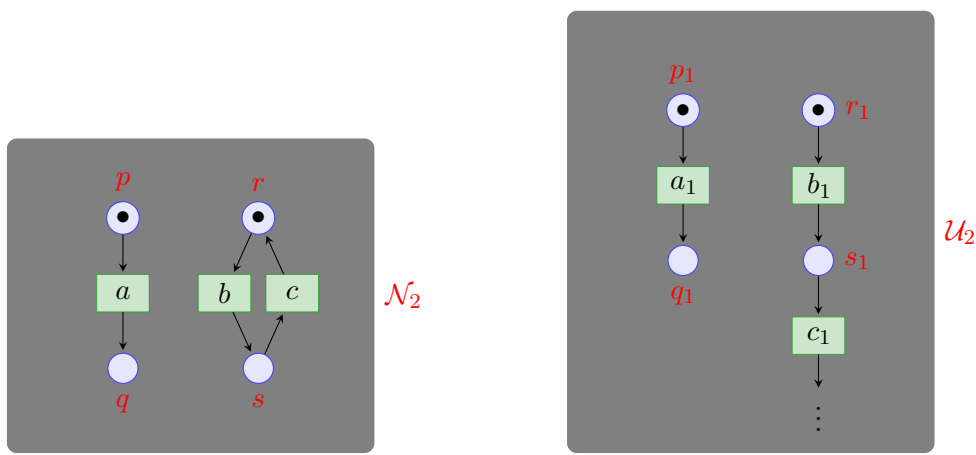
\begin{figure}[ht!]
  \centering
 \begin{subfigure}[t]{0.48\textwidth}  \def\a{1}
\def\b{0.45}
\def\c{1}
\begin{tikzpicture}[>=stealth,shorten >=1pt,node distance=\a cm,auto]
 \node[inner] at (-2*\b,-\c) (A1) [label = left:$\mathcal{N}_1$] {};  
 \node[outer] at (0,-2.3*\b) (A1) [label = right:$\mathcal{N}_2$] {};
  \node[cond] (p1) at (-2*\b, 0)  [label=above:\text{$p$}] {$\bullet$};
  \node[cond] (p2) at (-2*\b, -2*\c)  [label=below:\text{$q$}] {};
  \node[cond] (p3) at (2*\b, 0)  [label=above:\text{$r$}] {$\bullet$};
  \node[cond] (p4) at (2*\b, -2*\c)  [label=below:\text{$s$}] {};
  
  \node[event] (a) at (-2*\b, -\c) {$a$};
  \node[event] (b) at (\b, -\c) {$b$};
  \node[event] (c) at (3*\b, -\c) {$c$};
 \node[none,  left of = b](aux){};

 \path[->] (p1)  edge (a)
 (a) edge (p2)
       (p3) edge (b)
         (p4)  edge (c)
         (b) edge (p4)
        (c) edge (p3)
           ; 
\end{tikzpicture}
    \end{subfigure}
 \begin{subfigure}[t]{0.48\textwidth}  \def\a{1}
\def\b{0.45}
\def\c{1}
\begin{tikzpicture}[>=stealth,shorten >=1pt,node distance=\a cm,auto]
  \node[inner] at (-2*\b,-\c) (A1) [label = left:$\unfolding_1$] {};  
  \node[outer2] at (0,-3.6*\b) (A1) [label = right:$\unfolding_2$] {};
  \node[cond] (p1) at (-2*\b, 0)  [label=above:\text{$p_1$}] {$\bullet$};
  \node[cond] (p2) at (-2*\b, -2*\c)  [label=below:\text{$q_1$}] {};
  \node[cond] (p3) at (2*\b, 0)  [label=right:\text{$r_1$}] {$\bullet$};
  \node[cond] (p4) at (2*\b, -2*\c)  [label=right:\text{$s_1$}] {};
  
  \node[event] (a) at (-2*\b, -\c) {$a_1$};
  \node[event] (b) at (2*\b, -\c) {$b_1$};
  \node[event] (c) at (2*\b, -3*\c) {$c_1$};
  \node[none] (d) at (2*\b, -4*\c) {$\vdots$};
 \node[none,  left of = b](aux){};

 \path[->] (p1)  edge (a)
 (a) edge (p2)
       (p3) edge (b)
         (p4)  edge (c)
         (b) edge (p4)
        (c) edge (d)
           ; 
\end{tikzpicture}
    \end{subfigure}
\captionsetup{subrefformat=parens}
  \caption{\label{fig:ondooms} Left: Two safe Petri nets, $\petrinet_2$ properly contains $\petrinet_1$; right: their respective unfoldings  $\unfolding_1$ and $\unfolding_2$}
  \end{figure}
\paragraph{Remarks.} Some comments are in order here, since i) Definition~\ref{de:doom} introduces two different doomedness notions, and ii) clearly prefers the second over the first, in that the dual notion of freeness is defined without regard to \emph{strong doom}. To understand the motivation behind this choice, one must first 
appreciate the difference between doom and strong doom. Consider the toy example shown in Figure~\ref{fig:ondooms}, and suppose in both nets, a marking is bad iff it contains place $q$. Then the initial marking $\{p\}$ of $\petrinet_1$ is clearly both doomed and strongly doomed. However, marking $\markn_0=\{p,r\}$ and the configuration $\emptyset$ `leading to' $\markn_0$ in $\petrinet_2$ is  doomed, since the only maximal configuration of $\unfolding_2$ contains $a_1$. Nevertheless, $\emptyset$ is not strongly doomed, since with $\confign\define\{b_i,c_i:~i\in\NN\}$ we have $\emptyset\subsetneq\confign\in\infigs$, yet no finite configuration contained in $\confign$ is bad.

The difference in the two notions of doom thus lies in a semantic assumption of progress, or \emph{weak fairness}. While it is interesting in its own right, and potentially a subject for future work, to study the ramifications of the theory that build on the absence of progress, and hence on the notion of `\emph{strong doom}', we choose here to focus on the notion of doom given in (\ref{eq:doom}), and to consider those configurations that are \emph{doomed} to be configurations to avoid.
Once one assumes that the enabled transition $a$, that cannot be enabled, will fatally eventually fire, one is indeed led to seeing $\emptyset$ in Figure~\ref{fig:ondooms} as fatal, and as a state to avoid in any application.

Returning to the running example of Figures~\ref{fig:basinexample} and~\ref{fig:basinunfold}, if we consider $\badstates\define\{\{p_8\}\}$,  then configuration $\confign_1\define\{\alpha_1\}$ is free, and $\confign_2\define\confign_1\cup\{\gamma_1\}$ is doomed because $\confign_3\define\confign_2\cup\{\xi_1\}$ is bad.

By abuse of terminology, we will call any marking $\markn$ \emph{free} if there is a free $\confign\in\finfigs$ such that $\marking{\confign}=\markn$.
In the following,  we assume that the \emph{badness problem} `$\markn\in\badstates ?$' has been decided for every reachable marking $\markn$. In fact, a typical badness criteria can  be easily decided, e.g. by presence or absence of a fixed submarking. We will not dwell on the complexity of this decision problem as it is outside the scope of this article.

Since all bad markings are automatically doomed, we need to classify good markings into the  
doomed ones on the one hand, and the 
free ones on the other; the next two chapters will address this problem.

\section{Verification of Freeness}
In order to check whether a given marking $\markn$ is partially free, the key question is whether it is possible to reach, from $\markn$, some $\markn'\in\goodstates$ such that $\markn'$ admits a non-empty firing sequence $\sigma$ with $\markn'\move{\sigma}\markn'$. Indeed, under the assumption that $\petrinet$ is \emph{deadlock-free} in the sense that no maximal configuration contains any maximal event, it is easy to see that the existence of such an $\markn'$ is equivalent to partial freedom of $\markn$.
Of course, deadlock-freeness can always be obtained by adding dummy loop transitions to any deadlocked partial marking; the assumption therefore means no loss of generality.
Note that the classical algorithms for finding loops in a \emph{transition system} will not help us here, since the size of the state graph of most Petri nets we endeavour is prohibitive; we will adapt unfolding-based methods following~\cite{Esparza08}.
A procedure that, for any reachable marking $\markn$, generalizes that for McMillan's complete prefix from $\markn$, will produce a sufficient data structure for checking freedom of $\markn$.
\subsection{Search for loops}\label{sec:loop}
The key for obtaining this is obviously the power to identify loops in the reachability relation. We need to catch such loops at the earliest possible opportunity, in a small prefix.

\begin{definition}
    $\confign_1,\confign_2\in\finfigs$ are \emph{marking equivalent}, written $\confign_1\sim_\markn\confign_2$, iff $\marking{\confign_1}=\marking{\confign_2}$. Say that $\confign_1$ and $\confign_2$ form a \emph{loop}, written $\kringel{\confign_1,\confign_2}$, iff
    \begin{enumerate}
        \item they exhibit a cycle in the state graph, i.e. $\confign_1\sim_\markn\confign_2$ and $\confign_1\subsetneq\confign_2$, and 
        \item there are no configurations between $\confign_1$ and $\confign_2$ that exhibit such a cycle, i.e. there do not exist $\confign_3,\confign_4\in\finfigs$ such that 
        \begin{enumerate}
            \item $\confign_1\subseteq\confign_3\subsetneq\confign_4\subsetneq\confign_2$
\item $\confign_3\sim_\markn\confign_4$
\item $\{\confign_1,\confign_2\}\ne\{\confign_3,\confign_4\}$.
        \end{enumerate}
Write $\minkringel{\confign_1,\confign_2}$ iff $\confign_1$ and $\confign_2$ form a \emph{minimal loop} wrt inclusion, i.e. iff
\begin{enumerate}
    \item $\kringel{\confign_1,\confign_2}$
    \item for $\confign_1',\confign_2'\in\finfigs$, $\confign'_1\subsetneq\confign_1$ and $\confign'_2\subseteq\confign_2$ together imply that $\neg\kringel{\confign'_1,\confign'_2}$.
\end{enumerate}
    \end{enumerate}
\end{definition}
We have:
\begin{Lemma}
    Every loop-free configuration $\confign_\markn\in\finfigs$ such that
   $\marking{\confign_\markn}=\markn$ is in McMillan's (~\cite{McM92}) prefix $\Pi^{McM}_0(\markn)\define\Pi^{\subsetneq}_0(\markn)$.
\end{Lemma}
\bepr
Follows from the construction of $\Pi^{McM}_0(\markn)$ since configurations are only truncated by the cutoff criterion when they produce a loop. 
\eepr
\begin{Lemma}\label{le:contain}
    $\forall~\confign_1,\confign_2\in\finfigs(\markn_0)$: if $\minkringel{\confign_1,\confign_2}$, then $\confign_2$ is a configuration of $\Pi^{McM}_0(\markn)$ (and a fortiori, so is $\confign_1$).
\end{Lemma}
\bepr Suppose that under the assumptions of the lemma, $\confign_2$ is not a configuration of $\Pi^{McM}_0(\markn)$. In that case, it must contain $\confign_2^a\subseteq \confign_2^b\subsetneq \confign_2$ such that $\kringel{\confign_2^a,\confign_2^b}$. But $\minkringel{\confign_1,\confign_2}$ implies this is impossible unless $\confign_1=\confign_2^a$ and $\confign_2=\confign_2^b$. But if this is the smallest loop in $\confign_2$, then $\confign_2$ must be in $\Pi^{McM}_0(\markn)$.
\eepr
\begin{Lemma}
    For every  marking sequence $\markn\move{\transn_1}\ldots \markn$ that contains no properly smaller loop, 
    there is a configuration $\confign_\markn$ in $\Pi^{McM}_0(\markn)$ such that $\minkringel{\emptyset,\confign_\markn}$. 
\end{Lemma}
\bepr
A consequence of the definition of
$\Pi^{McM}_0(\markn)$
and of Lemma~\ref{le:contain}.
\eepr
\subsection{Verification of Freeness}
By the above discussion, 
inspection of $\Pi^{\subseteq}_0(\markn)$
yields all loops reachable from $\markn$; $\markn$ is free if any of them is good, i.e. such that the loop contains no bad marking. Equivalently, a loop $(\confign_1,\confign_2)$ is good iff $\marking{\confign_1}=\marking{\confign_2}\not\in\badstates$. Concretely, it suffices to consider every maximal event $e$ of $\Pi^{\subseteq}_0(\markn)$, since all these $e$ must have a mirror event $e'<e$ such that 
$\minkringel{\cone{e'},\cone{e}}$. If $\marking{e}$ is good, so is the entire loop.

This informal algorithm can be sped up by declaring, in addition to loop-cutoffs, any event $e$ such that $\marking{\cone{e}}\in\badstates$ (or equivalently, $\cone{e\in\badconfs}$) as a cutoff event in the unfolding procedure. To decide if $\marking{\cone{e}}\in\badstates$, we take advantage of $\badstates$'s reachability closure; all the initially known bad markings are used in $\netn$ (referred as the bad net, $\netn_{\bad}$) to unfold it so that we can check reachability and test whether a marking is bad or not, i.e., if $\marking{\cone{e}}\in\unfolding_{\netn_{\bad}}$ then $\marking{\cone{e}}$ is bad, and hence $e$ is a cutoff. The resulting prefix is slightly, sometimes considerably smaller than the full McMillan prefix; the worst case size of the latter remains, however, to be taken into account.
Below, this algorithm is assumed invoked by the boolean function \textsc{FreeCheck}$(\markn)$  that outputs \textsc{True} iff marking $\markn$ is free.

\section{Cliff-Edges and Ridges}

Recall that every reachable marking is represented by at least one configuration of the unfolding.
Moreover, since the future evolution of $\netn$ depends only on the current marking,  
$\marking{\confign_1}=\marking{\confign_2}$ for two configurations $\confign_1$ and $\confign_2$ implies that either both $\confign_1$ and $\confign_2$ are free, or both are doomed. 
Therefore, by extension, we call $\marking{\confign}$ free or doomed whenever $\confign$ is.

\textit{Running Example.} In the context of Figures~\ref{fig:basinexample} and~\ref{fig:basinunfold}, we consider $\badstates$ the singleton set containing the marking $\markn_8=\{\place_8\}$. Clearly, $\confign_1=\{\alpha_1,\gamma_1,\xi_1\}$ and $\confign_2=\{\beta_1,\delta_1,\eta_1\}$ satisfy $\marking{\confign_1}=\marking{\confign_2}=\markn_8$ and therefore $\confign_1,\confign_2\in\badconfs$. But note that $\confign_1'=\{\alpha_1,\gamma_1\}$ and $\confign_2'=\{\beta_1,\delta_1\}$ produce markings outside  $\badstates$, but they are doomed since any extension of these configurations leads into $\badstates$. Therefore, $\confign_1',\confign_2'\in\badconfs$. On the other hand,
$\emptyset$ is free, as well as $\{\beta_1,\gamma_1\}$, $\{\alpha_1,\delta_1\}$, etc.
We note in passing that the Petri net in Fig~\ref{fig:basinexample} allows to refine the understanding of the `tipping point' by showing that doom is not brought about by a single transition but rather the combined effect of two independent choices; this fact is obscured, or at least far from obvious, in the state graph shown in Figure~\ref{fig:basinstategr}.

Identifying free and doomed configurations belongs to the core objectives of this paper.

 From the minimal doomed configurations, we derive the critical `points' at which a run becomes doomed:
\begin{definition}
An event set $\watershed\subseteq\otrans$ is called a \emph{cliff-edge} iff there exists a minimally doomed configuration $\confign\in\mindofigs$ such that $\watershed=\crest{\confign}$. The set of cliff-edges is denoted $\tips$. The folding $\ridge\define\fold(\watershed)\subseteq\trans$ of a cliff-edge $\watershed$ is called a \emph{ridge}.
\end{definition}
To complete the map of the evolutional landscape for $\netn$, it is important to find, in a bounded prefix of the unfolding, all ridges that determine the viability of a trajectory. Notice that the completeness of prefix $\prefixn_0$ only guarantees that all reachable \emph{markings} of $\netn$ are represented by at least one configuration of $\prefixn_0$; this does not extend to a guarantee that all concurrent steps that lead into a doomed marking can be found in $\prefixn_0$ as well. Fortunately, one has: 
\begin{lem}
\label{le:ridge} For every ridge $\ridge$ of $\netn$ there is a witness in $\prefixn_0^{\subseteq}$, i.e. there exists a minimally doomed configuration $\confign$ in $\prefixn_0^{\subseteq}$ such that $\fold(\crest{\confign})=\ridge$.
\end{lem}
\bepr 
Fix $\ridge$, and let $\confign_\ridge$ be any configuration such that $\fold(\crest{\confign_\ridge})=\ridge$; set $\markn^\confign\define\marking{\confign_\ridge}$, and let $\markn_\ridge^\confign$ the unique reachable marking such that $\markn_\ridge^\confign\move{\ridge}\markn^\confign$. Then any such  $\markn_\ridge^\confign$ is represented by some $\confign^\ridge$ in $\prefixn_0^{\subseteq}$ by completeness.
\eepr

\section{Finding Minimally Doomed Configurations}
In the light of the above, we need to proceed in two steps. First, a configuration whose end events are without any immediate conflict - we will call such events \emph{unchallenged} - cannot be minimal; we will thus first describe how to \emph{shave} given configurations in order to approximate minimally doomed ones contained in them. Then we need to check whether a given configuration is doomed or free. The algorithm \textsc{MinDoo} will then combine both functions into a  search for minimally doomed configurations.
\subsection{Preparations: Shaving.} Let us start by observing that $\badconfs$, an upward closed set by construction, also has some downward closure properties, meaning one can restrict control to act on `small' configurations. The first idea is to remove maximal events $e$ from a configuration $\confign$ if they are not involved in any direct conflict; the idea is that in such a case, the reduced configuration
$\confign\backslash\{e\}$ has exactly the same maximal extensions as $\confign$. In fact,
$\confign\backslash\{e\}$ \emph{reveals} $e$ in a sense made precise in~\cite{BCH-fi12,Haar-tac10,HKS-tcs13,HRS-acsd13}; however, we will not be using exactly the relations introduced there.
\begin{definition}
{\it An event $\otransn$ is \emph{unchallenged} iff there is no $\otransn'$ such that $\otransn\dircf\otransn'$, i.e. $\postset{(\preset{\otransn})}=\{\otransn\}$. 
}
\end{definition}
\begin{lem}\label{le:shave}
Let $\confign\in\finfigs$ and $\otransn\in\crest{\confign}$ unchallenged; set $\confign'\define\confign\backslash\{\otransn\}$. Then $\confign'\in\finfigs$, and $\runs_\confign=\runs_{\confign'}$.
\end{lem}
\bepr 
$\confign'\in\finfigs$ holds by construction. Also, $\runs_{\confign}\subseteq\runs_{\confign'}$ follows from $\confign'\subseteq\confign$; it remains to show the reverse inclusion. Assume there exists $\runn\in\runs_{\confign'}\backslash\runs_\confign$; then $\confign\backslash\runn=\{\otransn\}$, and $\trunk{\otransn}\subseteq\runn$. By maximality, $\runn$ must contain some $\otransn'$ such that $\otransn\conflict\otransn'$. Then by definition, there are events $u\ne v$, $u\le e$, $v\le e'$, and $u\dircf v$.
In particular, $u\conflict e'$, and since $\{e'\}\cup\trunk{\otransn}\subseteq\runn$,
this implies $u=e$. But $e$ is unchallenged, so $v$ cannot exist, and neither can $\runn$.

\eepr

\begin{figure}
    \centering
 \def\a{1}
\def\b{0.95}
\def\c{0.85}
\begin{tikzpicture}[>=stealth,shorten >=1pt,node distance=\a cm,auto]
  \node[cond] (b1) at (0, 0)  [label=left:\text{$b_1$}] {};
  \node[cond] (b2) at (-1*\b, -2*\c)  [label=left:\text{$b_2$}] {};
  \node[cond] (b3) at (1*\b, -2*\c)  [label=right:\text{$b_3$}] {};
  \node[cond] (b4) at (-1*\b, -4*\c)  [label=right:\text{$b_4$}] {};
  \node[cond] (b5) at (1*\b, -4*\c)  [label=left:\text{$b_5$}] {};
  \node[cond] (b6) at (-3*\b, -6*\c)  [label=left:\text{$b_6$}] {};
  \node[cond] (b7) at (-1*\b, -6*\c)  [label=left:\text{$b_7$}] {};
  \node[cond] (b8) at (1*\b, -6*\c)  [label=right:\text{$b_8$}] {};
  \node[cond] (b9) at (3*\b, -6*\c)  [label=right:\text{$b_9$}] {};
 \node[cond] (b10) at (0*\b, -8*\c)  [label=left:\text{$b_{10}$}] {};

  \node[event] (x) at (0*\b, -\c) {$x$};
  \node[event] (y) at (-1*\b, -3*\c) {$y$};
  \node[event] (z) at (1*\b, -3*\c) {$z$};
\node[event] (a) at (-3*\b, -5*\c) {$\alpha$};
\node[event] (b) at (-1*\b, -5*\c) {$\beta$};
\node[event] (c) at (1*\b, -5*\c) {$\gamma$};
\node[event] (d) at (3*\b, -5*\c) {$\delta$};
\node[event] (u) at (0*\b, -7*\c) {$u$};

 \path[->] (b1)  edge (x)
       (x)  edge (b2)
      (x)  edge (b3)
      (b2) edge (y)
      (b3) edge (z)
      (y) edge (b4)
      (z) edge (b5)
      (b4) edge (a)
      (b4) edge (b)
      (b5) edge (c)
      (b5) edge (d)
      (a) edge (b6)
      (b) edge (b7)
      (c) edge (b8)
      (d) edge (b9)
      (b7) edge (u)
      (b8) edge (u)
      (u) edge (b10)
       ;
\end{tikzpicture}
    \caption{An occurrence net. With $\confign\define\{x,y,z,\beta,\gamma\}$ and $\confign'\define\confign\cup\{u\}$, suppose $\badstates=\{\marking{\confign'}\}=\fold(\{b_{10}\})$. Then  $\shaved{\confign'}=
    \confign$, and $\confign$ is doomed. Moreover, $\confign\in\mindofigs$ since both 
    $ \confign_3\define\confign\backslash\{\beta\}$ and $ \confign_4\define\confign\backslash\{\gamma\}$ are free.}
    \label{fig:wreath}
\end{figure}
\begin{definition}
A configuration $\confign\in\finfigs$ such that $\crest{\confign}$ contains no unchallenged event is called \emph{shaved}. 
\end{definition}
Clearly, every $\confign\in\finfigs$ contains a unique maximal shaved configuration, which we call $\shaved{\confign}$; it can be obtained from $\confign$ by recursively `shaving away' any unchallenged $\otransn\in\crest{\confign}$, and then continuing with the new crest, until no unchallenged events remain.

\textit{Example.} In the context of Figure~\ref{fig:wreath}, for $\confign_1=\{x,y,z\}$ and $\confign_2=\confign_1\cup\{\beta,\gamma,u\}$, one has $\shaved{\confign_1}=\emptyset$ since $x$, $y$, and $z$ are unchallenged, and $\shaved{\confign_2}=\confign_1\cup\{\beta,\gamma\}$ since $u$ is unchallenged but neither $\beta$ nor $\gamma$ are. 
Note that in the unfolding of the running example shown in Figure~\ref{fig:basinunfold}, the $\kappa$-labeled events are the only unchallenged ones.

As a consequence of Lemma~\ref{le:shave}, any $\confign\in\finfigs$ is in $\badconfs$ iff $\shaved{\confign}$ is. Still,  it may be possible that such a $\shaved{\confign}$ can still be reduced further by removing some of its crest events. This would be the case, e.g., if two conflicting events both lead to a bad state. Thus, given a crest event $e$, we test whether $\confign\backslash\{\otransn\}$ is free (e.g. because some event in conflict with $\otransn$ may allow to move away from doom) or still doomed. If the latter is the case, then $\confign$ was not minimally doomed, and analysis continues with $\confign\backslash\{\otransn\}$ (we say that we `rub away' $\otransn$). If $\confign\backslash\{\otransn\}$ is free, we leave $\otransn$ in place and test the remaining events from $\crest{\confign}$. A configuration that is shaved and from which no event can be rubbed away is minimally doomed. 

\subsection{Algorithm \textsc{MinDoo}}
Algorithm~\ref{alg:two} uses a `worklist' set $\worklist$ of doomed, shaved configurations to be explored; $\worklist$ is modified when a configuration is replaced by a set of rubbed (and again, shaved) versions of itself, or when a configuration $\confign$ is identified as minimally doomed, in which case it is removed from $\worklist$
 and added to $\mindooout$.
\begin{algorithm}[hbt!]
\caption{Algorithm \textsc{MinDoo}}\label{alg:two}
\KwData{Complete prefix $\prefixn_0$ of safe Petri Net $\petrinet=\tup{\place,\trans,\flow,\markn_0}$ and the set $\badconfs^0$  of bad configurations  of $\prefixn_0$}
\KwResult{The set $\mindooout$ of $\prefixn_0's$ $\subseteq$-minimal doomed configurations}
$\mindooout\gets\emptyset$;~
$\worklist \gets \emptyset$\;
\ForEach{$\confign\in\min_{\subseteq}(\badconfs^0)$}{
    $\confign'\gets\shaved{\confign}$\;
    $\worklist\gets\worklist\cup\{\confign'\}$\;
}
\While{$\worklist\neq\emptyset$}{
    Pick $\confign\in\worklist$;
    $\addit\gets \true$\;
    \uIf{NOT \textsc{FreeCheck}$(\marking{\confign\backslash\crest{\confign}})$}{
            $\addit\gets \false$\;
           $\confign'\gets\shaved{\confign\backslash\crest{\confign}}$\;
            $\worklist \gets (\worklist\cup\{\confign'\})$\;
            }
    \Else{
        \ForEach{$\otransn\in\crest{\confign}$}{
            \If{NOT \textsc{FreeCheck}$(\marking{(\confign\backslash\{\otransn\})}$}{
                $\addit\gets \false$\;
                $\confign'\gets\shaved{\confign\backslash\{\otransn\}}$\;
                $\worklist \gets \worklist\cup\{\confign'\}$\;
            }
    }}
    $\worklist\gets\worklist\backslash\{\confign\}$\;
    \If{$\addit$}{
        $\mindooout\gets\mindooout\cup\{\confign\}$\;
      }}
\Return{$\mindooout$}
\end{algorithm}

Every branch of \textsc{MinDoo} stops when a minimally doomed configuration is reached, i.e., a doomed configuration $\confign$ such by rubbing off any crest event $e$ from $\confign$ makes it free, i.e. $\confign\backslash\{e\}$ is free for all $e\in\crest{\confign}$. When the worklist is empty,  all minimally doomed configurations have been collected in~$\mindooout$.
Note that if $\emptyset\in\worklist$ at any stage during the execution of Algorithm~\textsc{Mindoo}, then
$\emptyset$ will be added to $\mindooout $, since \textsc{Mindoo} will not enter the second \textbf{foreach}-loop in that case.
In fact, if this situation arises, \emph{every} configuration is doomed, and thus  $\emptyset$ is the unique minimally doomed configuration.

The configurations produced in the course of the search strictly decrease w.r.t both size and inclusion. Moreover, an upper bound on the prefixes explored at each step is given by $\badconfs$, itself strictly contained in the complete finite prefix used to find all bad markings. 
According to~\cite{ERV02}, this prefix can be chosen of size equal or smaller (typically: considerably smaller) than the reachability graph of $\petrinet$ (number of non-cut-off events in the prefix are less or equal than the number of reachable markings in $\petrinet$).

\begin{thm}{\it
For any safe Petri net $\petrinet=\tup{\net,\markn_0}$ and bad states set $\badstates\subseteq \reach_\net{(\markn_0)}$, Algorithm \textsc{MinDoo} terminates, with output set $\mindooout$ containing exactly all minimal doomed configurations, i.e. $\mindooout=\mindofigs$.}
\end{thm}
\bepr 
Termination follows from the finiteness of $\min_\subseteq(\badconfs_0)$,
since in  each round of \textsc{MinDoo} there is one configuration $\confign$ that is either replaced by a set of strict prefixes or removed from $\worklist$. Therefore, after a finite number of steps, $\worklist$ is 
empty. 

As shown in Section~\ref{sec:loop}, the status (doomed or free) of a given finite configuration can effectively be checked on a fixed finite prefix of $\unfolding$.
Assume that after termination of \textsc{MinDoo}, one has $\confign\in\mindooout$; we need to show $\confign\in\mindofigs$. Clearly, when $\confign$ was added to $\mindooout$, it had been detected as doomed; it remains to show that $\confign$ is also minimal with this property. Assume that there is $\confign'\subsetneq\confign$ that is doomed as well. But in that case  there exists $\otransn\in\crest{\confign}$  such that $\confign'\subseteq(\confign\backslash\{\otransn\})\subsetneq\confign$, which implies that this  $(\confign\backslash\{\otransn\})$ is doomed as well. But then $\addit$ has been set to $\false$ in the second \textbf{foreach}-loop, before $\confign$ could have been added to $\mindooout$. 

Conversely, let $\confign\in\mindofigs$. Then $(\confign\backslash\{\otransn\})$ is free for all $\otransn\in\crest{\confign}$; the variable $\addit$ remains thus at the value $\true$ because no round of the second \textbf{foreach}-loop can flip it. Thus $\confign$ is added to $\mindooout$, from which \textsc{MinDoo} never removes any configuration.

\eepr 
\ifexperiments

\subsection{Implementation and Experiments.}
An implementation of \textsc{Mindoo} is available at~\cite{ecofolder2024} using the module \textsc{doomed}.
It takes as input a safe Petri net in the PEP~\cite{PEP} format, the list of undesired (or bad) markings separated by newlines and the unfolding of the bad net, it relies on \textsc{Ecofolder}~\cite{ecofolder2024} for computing the
complete finite prefix $\prefixn_0$ of both the system and the bad net. Also, \textsc{Ecofolder} is used to apply McMillan's criterion~\cite{McM92} to unfold every corresponding system's net initialized with a marking of a configuration's crest choosen from $\worklist$. In the unfolding process,
we perform reachability checks using Answer-Set programming (ASP) to know whether the marking is reachable in the bad net, then decide its freeness status. 
We have  implemented Algorithm~\ref{alg:two} in Python, using ASP for the identification of bad configurations, specifically  with the help of 
 the \textsc{Clingo} solver ~\cite{clingo}.

 Table~\ref{tab:experiments} illustrates the performance of the implementation on different
instances of Petri nets modeling biological processes.
\begin{table}
    \caption{Statistics of Algorithm~\ref{alg:two} on Petri net models of biological systems.
        $|\place|$ is the number of places in the system; $|\trans|$ is the number of transitions; the size of $\prefixn_0$ is the number of their events;
        $|\badstates|$ is the number of bad markings; $|\worklist|$ is the number of bad configurations initially identified;
        ``\# free checks'' is the number of freeness checks for \textit{free} status of a configuration.
    ``time'' is the total computation time on a 1.8Ghz CPU.
    \label{tab:experiments}}
\begin{tabular}{|l|c|c|c|c|c|c|c|c|}\hline \rule{0pt}{5mm}
Model & $|\place|$ & $|\trans|$ & size $\prefixn_0$ & $|\badstates|$ & $|\worklist|$ & $|\mindofigs|$ & \# free checks & time\\[1mm]\hline
Lambda switch & 11 & 41 & 126 & 1 & 5 & 5 & 13 & 1s\\
Mammalian cell cycle & 20 & 38 & 176 & 1 & 25 & 0 & 78 & 1s \\
Cell death receptor & 22 & 33 & 791 & 1 & 16 & 14 & 146 & 9m57s \\
Budding yeast cell cycle & 18 & 32 & 1,413 & 1 & 30 & 28 & 165 & 2m28s\\
\hline
\end{tabular}
\end{table}
In each case, we report the number of places and transitions, the size (number of events) of prefix $\prefixn_0$ (including cut-off events), the number of \textit{bad} markings initially given to the algorithm ($|\badstates|$), the number of bad configurations in $\worklist$ leading to the given bad markings, the number of minimally doomed configurations ($|\mindofigs|$), the number of
configurations which have been tested for being free, and the total time.
The aim of the experiments conducted in this study was to investigate the feasibility and effectiveness of our approach for analyzing standard  models of biological systems from the literature. Specifically, we focused on models for which the study of doomed configurations was relevant, as these configurations can provide important insights into the behavior and properties of the system.
Potential bottlenecks include the computation of those maximal configurations that lead  to a \textit{bad} marking, and most importantly the unfolding process using McMillan's cutoff criterion~\cite{mcmillan1993} for the freeness test since, as shown in~\cite{ERV02}, it can be exponentially larger than the number of reachable markings of the net. Hereafter, our experiments focused on evaluating the impact of different number of places and transitions, and prefix sizes ($\prefixn_0$) on several metrics, including the number of minimally doomed configurations, the number of candidate configurations screened by Algorithm~\ref{alg:two} (\# of free checks in Table~\ref{tab:experiments}), and the overall computation time as 
 the prefix size and the number of nodes in the net increased. 
 We have selected four models that had been initially published as Boolean networks, which can be translated into equivalent safe Petri nets using the encoding described in \cite{CHJPS-cmsb14} and implemented in the tool \textsc{Pint}~\cite{Pint-CMSB17}.

The ``Lambda switch" model~\cite{tt95} comprises 11 places and 41 transitions, and is a gene regulatory network for a bacterial virus known as the lambda phage. This virus is a type of temperate bacteriophage, which means it can establish a long-term symbiotic relationship with its bacterial host, known as the \textit{lysogenic} response. In this state, the virus is faithfully transmitted to the bacterial progeny. However, in most cases, the virus follows the \textit{lytic} response, where it replicates itself, destroys the host cell, and ultimately lyses the cell. This dichotomous decision is cell-dependent and is regulated by intertwined feedback mechanisms involving four genes: cI, cro, cII, and N. In our experiments, we define the lytic response as ``bad" or undesired, and initialize the cro gene with a token while leaving all other genes without tokens.

The ``Mammalian cell cycle" model~\cite{faure2006} compromises 20 places and 38 transitions, reproducing the main known dynamical features of the wild-type biological system. Mammalian cell division is a highly regulated process that must be coordinated with the overall growth and development of the organism. This coordination is necessary to ensure that cell division occurs only when needed, such as during tissue repair or in response to hormonal signals. The decision of whether a cell will divide or remain in a resting state (known as quiescence or G0 phase) is determined by a complex interplay of extracellular positive and negative signals. These signals can include growth factors, cytokines, and other signaling molecules that bind to specific receptors on the cell surface. The balance of these signals ultimately determines whether the cell will enter the cell cycle or arrest the process; one of the factors causing cell arrest is disruption on Cyclin D (CycD) and its associated CDKs since they are vital for the transition from G1 to S phase. Therefore, we determine CycD disruption as a bad marking resulting in an empty set of minimally doomed configurations. In other words, the only configuration in the prefix that can avoid cell arrest is the empty set, a  finding that is suggestive in itself.

The ``Cell death receptor'' model~\cite{calzone2010} comprises 22 places and 33 transitions, and reproduces a bifurcation process into different cell fates, one of which has been declared as bad (apoptosis). The model focuses on the activation of death receptors (TNF and FAS) in various cell types and conditions. The cell's fate can vary significantly as the same signal can trigger survival by activating the NFkB signaling pathway or lead to death by apoptosis or necrosis. The study reveals the complex interplay and mutual inhibition between the NFkB pro-survival, RIP1-dependent necrosis, and apoptosis pathways. Our analysis shows that the minimally doomed configurations identify the configurations in which a decisive event has occurred, committing the system to the undesirable attractor marked as bad. By detecting these configurations, biologists  gain insight into the causal steps that lead to cell death via apoptosis, and identify decisive points where alternative pathways are still possible.

The ``Budding yeast cell cycle" model~\cite{Orlando2008} comprises 18 places and 32 transitions, capturing the oscillatory behavior of gene activity throughout the cell cycle. The biochemical oscillator controlling periodic events during the cell cycle is centered on the activity of cyclin-dependent kinases (CDKs), which are thought to play a crucial role in controlling the temporally ordered program of transcription in somatic cells and yeast. However, the study~\cite{Orlando2008} had integrated genome-wide transcription data and built models, in which periodic transcription emerges as a property of a transcription factor network. The authors investigated the dynamics of genome-wide transcription in budding yeast cells disrupted for all S-phase and mitotic cyclins to determine the extent to which CDKs and transcription factor networks contribute to global regulation of the cell-cycle transcription program. In our analysis, we use this model to identify the minimally doomed configurations that lead to a bad marking, in which the cycle exits its oscillatory behavior and all genes become inactive. By detecting these configurations, we can precisely identify the underlying factors that lead to the system exiting its oscillatory behavior.

\fi

 In each case, the number of minimally doomed configurations is a fraction of the size of the finite complete prefix $\prefixn_0$. The computation time for identifying minimally doomed configurations is primarily affected, as previously mentioned, by the potential bottleneck of checking for loops using McMillan's cutoff criterion. 
 In general, as the number of places increases, it naturally becomes more difficult to determine whether a marking is reachable from a bad one.  However, the particulary of the freeness check lies in the fact that one needs to unfold McMillan's prefix in order to be sure to have detected all cases in which a loop exists. In the worst case, this prefix may be  exponentially larger than the reachability graph; and in any case, one has to explore individual branches to a considerable depth, leading in turn to exponential branching in the prefix constructed. 
 
 Future work may explore ways to handle this bottleneck more efficiently, by accelerating computation speed for loop checks, particularly in hard and complex examples. 
 
 Another relevant challenge is to find compact representations of the set of minimally doomed configurations; indeed, as these configurations often share many events, biological interpretation may be facilitated by  regrouping them in a helpful way.

\section{Protectedness}

\subsection{Measuring the Distance from Doom}
\subsubsection*{Decisional Height.}
With the above, we have the tools to draw a map of the `landscape' in which the system evolves, with doomed zones and cliff-edges highlighted. What we wish to add now is to assist \emph{navigation} in this landscape: we intend to give a meaningful measure of how well, or badly, a current system state is protected against falling from a cliff-edge, that is, how far the system  is from entering a doomed state. We have chosen to measure this distance in terms not  of the \emph{length} of paths, or of similar notions, but rather in terms of the \emph{choices} that are made by the system in following  a particular path.

Consider a  configuration $\confign$ and the non-sequential process that it represents. Some of the events in $\confign$ can be seen as representing a \emph{decision}, in the sense that their occurrence took place in  conflict with some event that was enabled by some prefix of  $\confign$. The number of such events gives a measure of the information contained in $\confign$, in terms of the decisions necessary to obtain $\confign$:
\begin{definition}
Let $\confign\in\finfigs$, and define
\begin{eqnarray*}\label{eq:decision}
\dheight{\confign}&\define&\left|\left\{\otransn\in\confign:~\exists~\otransn'\in\otrans:~
\otransn\strcf^\confign\otransn'
\right\}\right|,
\end{eqnarray*}
where $\strcf^\confign$ is the \emph{strict $\confign$-conflict} relation defined, for all $\otransn\in\confign$, by
\begin{eqnarray*}
\otransn\strcf^\confign\otransn'&\setgdw&\otransn\dircf\otransn'~\land~\trunk{\otransn'}\subseteq\confign.
\end{eqnarray*}

$\dheight{\confign}$ is  called the \emph{decisional height} of $\confign$.
\end{definition}
In Figure~\ref{fig:basinunfold}, the configuration $\confign_1=\{\xi_1,\alpha_1,\gamma_1\}$  satisfies $\dheight{\confign_1}=2$, whereas for $\confign_0=\{\beta_1\}$, one has $\dheight{\confign_0}=1$.

\begin{figure}
    \centering
    \def\a{1}
\def\b{0.95}
\def\c{0.85}
\begin{tikzpicture}[>=stealth,shorten >=1pt,node distance=\a cm,auto]
  \node[cond] (b1) at (-1*\b0, 0)  [label=above:\text{$b_1$}] {};
  \node[cond] (b2) at (1*\b,0)  [label=above:\text{$b_2$}] {};
  \node[cond] (b3) at (-\b, -2*\c)  [label=left:\text{$b_3$}] {};
  \node[cond] (b4) at (1*\b, -2*\c)  [label=left:\text{$b_4$}] {};
  \node[cond] (b5) at (3*\b, -2*\c)  [label=right:\text{$b_5$}] {};
  \node[cond] (b6) at (-2*\b, -4*\c)  [label=left:\text{$b_6$}] {};
  \node[cond] (b7) at (0*\b, -4*\c)  [label=left:\text{$b_7$}] {};
  \node[cond] (b8) at (2*\b, -4*\c)  [label=right:\text{$b_8$}] {};

  \node[event] (x) at (-\b, -\c) {$x$};
  \node[event] (y) at (1*\b, -\c) {$y$};
  \node[event] (z) at (3*\b, -\c) {$z$};
\node[event] (a) at (-2*\b, -3*\c) {$\alpha$};
\node[event] (b) at (-0*\b, -3*\c) {$\beta$};
\node[event] (c) at (2*\b, -3*\c) {$\gamma$};

 \path[->] (b1)  edge (x)
        (x)  edge (b3)
      (b2) edge (y)
      (b2) edge (z)
       (y) edge (b4)
      (z) edge (b5)
      (b3) edge (a)
   (b3) edge (b)
      (b4) edge (b)
      (b4) edge (c)
      (a) edge (b6)
      (b) edge (b7)
      (c) edge (b8)
       ;
\end{tikzpicture}
    \caption{Illustration of direct conflict.}
    \label{fig:onconflicts}
\end{figure}
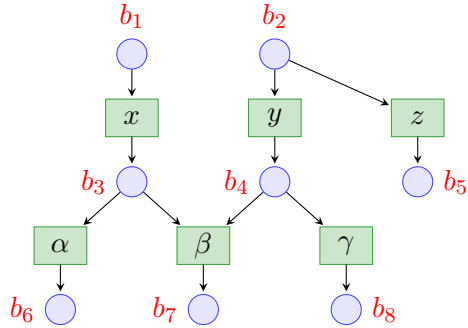

Figure~\ref{fig:onconflicts} shows an occurrence net with configuration $\confign_\beta\define\{x,y,\beta\}$ that has $\dheight{\confign_\beta}=3$ (because of $z,\alpha$, and $\gamma$) and configuration $\confign_\alpha\define\{x,z,\alpha\}$ with $\dheight{\confign_\alpha}=1$ (because of $y$). 

Note that $\strcf^\confign$ is more restrictive than direct conflict $\dircf$; it is also more restrictive than  the 
\emph{immediate conflict} in the literature (e.g.~\cite{abbes:hal-00350226}). It is closely dependent on the configuration $\confign$ under study, and describes precisely those events \emph{against} which the process had to decide in performing $\confign$.

\begin{figure}
    \centering
    \def\a{1}
\def\b{0.95}
\def\c{0.85}
\begin{tikzpicture}[>=stealth,shorten >=1pt,node distance=\a cm,auto]
  \node[cond] (b1) at (-1*\b, 0)  [label=left:\text{$b_1$}] {};
 \node[cond] (b2) at (1*\b, 0)  [label=left:\text{$b_2$}] {};
  \node[cond] (b3) at (-1*\b, -2*\c)  [label=left:\text{$b_3$}] {};
  \node[cond] (b4) at (1*\b, -2*\c)  [label=right:\text{$b_4$}] {};
  \node[cond] (b5) at (-2*\b, -4*\c)  [label=right:\text{$b_5$}] {};
  \node[cond] (b6) at (0, -4*\c)  [label=left:\text{$b_6$}] {};
  \node[cond] (b7) at (2*\b, -4*\c)  [label=left:\text{$b_7$}] {};

  \node[event] (x) at (-1*\b, -\c) {$x$};
  \node[event] (y) at (1*\b, -\c) {$y$};

\node[event] (a) at (-2*\b, -3*\c) {$\alpha$};
\node[event] (b) at (0, -3*\c) {$\beta$};
\node[event] (c) at (2*\b, -3*\c) {$\gamma$};

 \path[->] (b1)  edge (x)
      (b2)  edge (y)
       (x)  edge (b3)
      (y)  edge (b4)
      (b3) edge (a)
      (b3) edge (b)
      (b4) edge (b)
      (b4) edge (c)
      (a) edge (b5)
      (b) edge (b6)
      (c) edge (b7)
       ;
\end{tikzpicture}
    \caption{An `unprotected' branching process}
    \label{fig:nonseqdoom}
\end{figure}
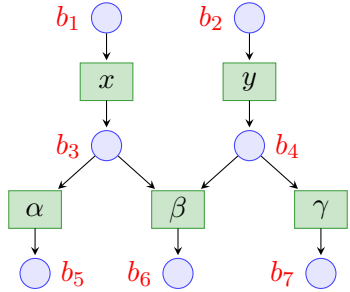

One may wonder why we choose this particular definition of \emph{decision}, rather than using the customary direct or immediate conflicts. The reason is that we wish to consider as decisions only deliberate actions against or in favor particular branches in a bifurcation situation, and \emph{not} any resolution of conflicts brought about by the nondeterminism in the `race' between two concurrent processes. Consider Figure~\ref{fig:nonseqdoom}, assuming that the only bad configuration is $\confign_\badconfs=\{\alpha,\gamma,x,y\}$. Then
clearly, the configurations $\confign_\alpha\define\{\alpha,x\}$ and  $\confign_\gamma\define\{\gamma,y\}$ are both doomed. Their height, measured by either direct or immediate conflict, would be $1$; in our definition, it is $0$, since $\alpha$ (or $\gamma$, respectively) was enabled  by $\confign_\alpha\backslash\{\alpha\}$ (or $\confign_\gamma\backslash\{\gamma\}$) without any conflict, since neither $\confign_\alpha\backslash\{\alpha\}$ nor $\confign_\gamma\backslash\{\gamma\}$ enabled $\beta$. Underlying this is the fact that the `decision' against $\beta$ is taken here without any choice, merely by the fact that of the two concurrent events $x$ and $y$, one may occur much faster than the other, creating a situation in which $\alpha$ (or $\gamma$) has no competitor.

\subsubsection*{Defining Protectedness.}
Now, for any free marking $\markn$ (or, equivalently, any free configuration $\confign$ such that $\marking{\confign}=\markn$),
we wish to measure the threat represented by doomed markings reachable from $\markn$: how far away from doom is the system when it is in $\markn$ ? Using the decisional height introduced above, we can define a height difference in terms of the conflicts that lead from one marking to another: 
\begin{definition}
For $\confign\in\finfigs$, let 
\begin{eqnarray}
\mindofigs_\confign&\define & \left\{ \begin{array}{lcr}
\{\confign'\in\mindofigs:~\confign\subseteq\confign'\}&: & 
\confign\in\viables\\
    \{\confign\} &:& \confign\in\badconfs
\end{array}\right.
\end{eqnarray}
The \emph{protectedness} of $\confign$ is then
\begin{eqnarray}
\schutz{\confign}&\define& \min_{\confign'\in\mindofigs_\confign}\left\{\dheight{\confign'
\backslash\confign}\right\}
 \end{eqnarray}
\end{definition}
 In Figure~\ref{fig:wreath}, with the definitions introduced there, $\schutz{\confign}=\schutz{\confign'}=0$. Setting $\confign_1\define\{x\}$, $\confign_2\define\{x,y\}$, $\confign_3\define\{x,z\}$, $\confign_4\define\confign_2\cup\confign_3$, $\confign_5\define\confign_4\cup\{\beta\}$,
$\confign_6\define\confign_4\cup\{\gamma\}$,
one further has
\begin{eqnarray*}
\schutz{\confign_1}=\schutz{\confign_2}=\schutz{\confign_3}=\schutz{\confign_4}&=&2\\
\schutz{\confign_5}=\schutz{\confign_6}&=&1.
\end{eqnarray*}
Returning to Figure~\ref{fig:onconflicts}, suppose that $\confign'=\{x,y,\beta\}$ is the only minimally doomed configuration. Then for $\confign=\{x,z,\alpha\}$ as above, we have $\schutz{\confign}=1$, because the only strict (and direct) conflict here is the one between $z$ and $y$. 

Note that the definition of protectedness is parametrized by the choice of conflict relation in computing $\dheight{\bullet}$. Using direct conflict instead of strict conflict would increase $\dheight{\bullet}$ and lead to an overevaluation of protectedness.

To see the point, consider the occurrence net in Figure~\ref{fig:onconflicts}. Let $\confign_\alpha=\{x,z,\alpha\}$, $\confign_\beta=\{x,y,\beta\}$  and $\confign_\gamma=\{x,y,\alpha,\gamma\}$. We have
$\dheight{\confign_\alpha}=1$, $\dheight{\confign_\beta}=3$ and $\dheight{\confign_\gamma}=2$. Were $\strcf$ replaced by $\dircf$ in the computation of $\dheight{\bullet}$, these values would not change \emph{except} for $\confign_\alpha$ where it would change to $2$. As a result, if $\confign\in\mindofigs$, the protectedness of the empty configuration would be evaluated as $2$, whereas by our definition $\schutz{\emptyset}=1$. Indeed, $\emptyset$ is \emph{just one wrong decision away from doom}, and this is what protectness is meant to express.

An even starker illustration is once again provided by Figure~\ref{fig:nonseqdoom}: in fact, we have $\schutz{\emptyset}=0$, as follows from the discussion above.
\subsection{Computing Protectedness is Feasible}
Computation of $\schutz{\bullet}$ does not require any larger data structure than those already required for  computing $\mindofigs$. In fact, an alternative - and often  much smaller prefix - is also sufficient:
\begin{lem}
\label{fig:schutz} There exists an adequate total order $\adequate$  such that the associate complete prefix scheme  producing  $\prefixn^\adequate_0$  whose size is bounded by the number of reachable markings, and such that for every finite configuration $\confign,$ $\schutz{\confign}$ can be computed on $\prefixn_0^\adequate(\marking{\confign})$.
\end{lem}
\bepr If $\mindofigs\cap\configs(\prefixn_0)=\emptyset$, then all extensions of $\confign$ are free, and we are done. Otherwise, we have to find an adequate \emph{total} order $\adequate$ on finite configurations, that ensures that $\prefixn^\adequate_0$ contains at least one  minimally doomed configuration that  minimizes $\dheight{\bullet}$ over all minimally doomed configurations in  $\unfolding(\marking{\confign})$.
The following order $\prec$ is obtained by modifying the total order $\prec_F$ introduced in~\cite{ERV02}, Def. 6.2.:
For $\confign_1,\confign_2\in\finfigs$, write $\confign_1\adequate\confign_2$ iff either 
\begin{itemize}
    \item $\dheight{\confign_1}<\dheight{\confign_2}$, or
    \item $\dheight{\confign_1}=\dheight{\confign_2}$ and $\confign_1\ll\confign_2$, or
    \item $\dheight{\confign_1}=\dheight{\confign_2}$ and $\confign_1\equiv\confign_2$, and $FC(\confign_1)\ll FC(\confign_2)$,
\end{itemize}
where $\ll$ ($\equiv$) denote lexicographic ordering (lexicographic equivalence) wrt some total ordering of the transition set $\trans$, and $FC$ denotes Cartier-Foata normal form. The proof of Theorem 6.4. of~\cite{ERV02} extends immediately, proving that $\prec$ is an adequate total order; therefore, Lemma 5.3. of~\cite{ERV02} applies, hence any complete prefix $\prefixn_0^\prec$ obtained via the scheme using $\prec$ is bounded in size by the reachability graph. Now, let $\configs^*$ be the set of configurations from $\mindofigs(\marking{\confign})$ that minimize $\dheight{\bullet}$; by construction of $\prec$, one has $\configs^*\cap\configs(\prefixn_0^\prec)\ne\emptyset$.
\eepr
\section{Discussion}
The results presented here contain, extend and complete those in our conference paper~\cite{Gandalf22}. The  toolkit for the analysis of tipping situations in a safe Petri net, i.e. when and how a basin boundary is crossed; an algorithmic method for finding minimally doomed configuration has been developed, implemented and tested. 

Moreover, we have introduced a measure of \emph{protectedness} that indicates the number of \emph{decisions} that separate a free state from doom. It uses an intrinsic notion of decisional height that allows to warn about impending dangerous scenarios; at the same time, this height is also `natural' for unfoldings, in the sense that it induces an adequate linear order that allows to compute complete prefixes of bounded size.

On a more general level, the results here are part of a broader effort to provide a discrete, Petri-net based framework for dynamical systems analysis in the life sciences. The applications that we target here lie in systems biology and ecology. 

Future work will investigate 
possibilities for  \emph{Doom Avoidance Control}, i.e. devising strategies that allow to steer away from doom; we expect to complement the existing approaches via structural methods of e.g. Antsaklis et al \cite{IorAnt2003,IorAnt2006}, and also the unfolding construction of Giua and Xie~\cite{GiuaXie2005}.  A crucial question is the knowledge that any control player can be assumed to have, as a basis for chosing control actions. We believe the protectedness measure is a valid candidate for coding this information, so that a controller may take action when the system is too close to doom (wrt some thresholds to be calibrated) but there still remain decisions that can be taken to avoid it. 
Evaluating this option, along with other approaches, must, however, be left to future work.

\textbf{Acknowlegments:}
We gratefully acknowledge the fruitful exchanges with C\'edric Gaucherel and Franck Pommereau. This work was supported by the \emph{DIGICOSME} grant \textsc{Escape}, \emph{DIGICOSME RD 242-ESCAPE-15203}, and by the French Agence Nationale pour la Recherche (ANR) in the scope of the
project ``BNeDiction'' (grant number ANR-20-CE45-0001).

\bibliographystyle{alphaurl}
\bibliography{output}  

\appendix
                            
\end{document}